# Chiral Polarization Textures Induced by the Flexoelectric Effect in Ferroelectric Nanocylinders

Anna N. Morozovska [1*], Riccardo Hertel [2†], Salia Cherifi-Hertel [2], Victor Yu. Reshetnyak [3], Eugene A. Eliseev [4], and Dean R. Evans [5‡]

[1] Institute of Physics, National Academy of Sciences of Ukraine,
46, pr. Nauky, 03028 Kyiv, Ukraine

[2] Université de Strasbourg, CNRS, Institut de Physique et Chimie des Matériaux de Strasbourg,
UMR 7504, 67000 Strasbourg, France

[3] Taras Shevchenko National University of Kyiv,
Volodymyrska Street 64, Kyiv, 01601, Ukraine

[4] Institute for Problems of Materials Science, National Academy of Sciences of Ukraine,
Krjijanovskogo 3, 03142 Kyiv, Ukraine

[5] Air Force Research Laboratory, Materials and Manufacturing Directorate,
Wright-Patterson Air Force Base, Ohio, 45433, USA

## Abstract

Polar chiral structures have recently attracted much interest within the scientific community, as they pave the way towards innovative device concepts similar to the developments achieved in nanomagnetism. Despite the growing interest, many fundamental questions related to the mechanisms controlling the appearance and stability of ferroelectric topological structures remain open. In this context, ferroelectric nanoparticles provide a flexible playground for such investigations. Here, we present a theoretical study of ferroelectric polar textures in a cylindrical core-shell nanoparticle. The calculations reveal a chiral polarization structure containing two oppositely oriented diffuse axial domains located near the cylinder ends, separated by a region with a zero-axial polarization. We name this polarization configuration "flexon" to underline the flexoelectric nature of its axial polarization. Analytical calculations and numerical simulation results show that the flexon's chirality can be switched by reversing the sign of the flexoelectric coefficient. Furthermore, the anisotropy of the flexoelectric coupling is found to critically influence the polarization texture and domain morphology. The flexon rounded shape, combined with its distinct chiral properties and the localization nature near the surface, are reminiscent of Chiral Bobber structures in magnetism. In the azimuthal plane, the flexon displays the polarization state of a vortex with an axially polarized core region, i.e., a meron. The flexoelectric effect, which couples the electric polarization and elastic strain gradients, plays a determining role in the stabilization of these chiral states. We discuss similarities between this interaction and the recently predicted ferroelectric Dyzaloshinskii-Moriya interaction leading to chiral polarization states.

---

\*      Corresponding author 1: anna.n.morozovska@gmail.com
†      Corresponding author 2: riccardo.hertel@ipcms.unistra.fr
‡      Corresponding author 4: dean.evans@afresearchlab.com

## I. INTRODUCTION

Research on ferroelectric materials has received growing interest over the past years, driven in part by the potential of these material systems for low-power technological applications in a broad spectrum of domains [1, 2], ranging from high-density data storage to optical nano-devices. A central aspect of this field of research is the formation of ferroelectric domain structures [3], and more generally the micro- and nanoscale structure of the polarization field [4]. Traditionally, research on ferroelectrics is centered on the study of bulk materials and thin films [5, 6, 7], but recently ferroelectric nanoparticles have also attracted increasing interest [8, 9, 10, 11, 12, 13, 14]. In ferroelectric thin films and nanoparticles, the polarization structure is strongly affected by electrostatic (depolarizing) fields [15, 16, 17, 18], as well as by strain and strain gradients [19, 20, 21, 22] via the flexoelectric effect [23, 24, 25, 26].

Although the foundations for the theoretical description of ferroelectrics have been established decades ago [27], understanding the complex physical properties of these material systems remains a challenge for fundamental research. Recent progress in this field, achieved to a large extent through advanced imaging techniques [28] and by employing modern numerical simulations [29], includes the discovery of highly complex polarization structures, such as flux closure [5, 30] and bubble domains [31], meandering [32, 33] and/or labyrinthine [11, 34] structures, non-Ising type chiral domain walls [35], polarization vortices in thin layers [36, 37, 38], nanodots [39] or nanopillars [40], or polar skyrmions [41, 42].

While skyrmions and other chiral structures have dominated the past decade of research in magnetism [43], these topological states have received less attention by the ferroelectric community. Only recently a strong interest has emerged in chiral polarization structures, which can be attributed to the observation of skyrmion states in ferroelectrics [41-42]. However, the theoretical understanding of these structures is not as advanced as it is in the case of their magnetic counterparts, and the mechanism that underpins the formation of skyrmions in ferroelectrics is not fully understood. The fundamental interaction stabilizing the magnetic version of these structures in chiral ferromagnets [44, 45] is the Dzyaloshinky-Moriya Interaction (**DMI**). The DMI favors the formation of helical structures with a well-defined handedness as they occur, e.g., along the radial direction of skyrmions. As scientists working on ferroelectrics hope to replicate the success that chiral structures have witnessed in magnetism, the possibility of a "ferroelectric DMI" has recently been discussed [46]. However, Erb and Hlinka [47] showed that only very few exotic ferroelectrics could theoretically sustain an intrinsic DMI-type interaction since it requires particular symmetry properties of the crystal lattice. Here we discuss the flexoelectric coupling as an alternative mechanism that can generate chiral polarization states in ferroelectrics.

The thermodynamic description of the flexoelectric effect is given by the Lifshitz invariant in the free energy expansion [22]. It is known that, in magnetic materials, the occurrence of similar Lifshitz invariants converts directly into an antisymmetric coupling known as the DMI [48, 49], which favors the formation of helicoidal structures with a specific chirality. The existence of a ferroelectric counterpart of the DMI was recently predicted by first-principles simulations [46]. The ferroelectric analogue of the DMI was discussed in the context of Lifshitz invariants by Strukov and Levanyuk [50], and more recently by Erb and Hlinka [47], who argued that a ferroelectric DMI can exist. In addition to the remarkable similarity in the mathematical form of the flexoelectric Lifshitz invariant and DMI, the flexoelectric term appears to have a similar impact as the DMI in terms of the formation of chiral structures.

By means of the finite element modeling (**FEM**) based on the Landau-Ginzburg-Devonshire (**LGD**) theory, this paper shows that an anisotropic flexoelectric effect can give rise to a previously unexplored type of polarization state with distinct chiral properties. Remarkably, these homochiral properties are not induced by a DMI term. This finding suggests that the recently discussed DMI in ferroelectrics is not the only possible mechanism for the formation of homochiral polarization states, and that anisotropic flexoelectric effects offer an alternative pathway to stabilize such structures in ferroelectric nanostructures. We discuss common aspects of the DMI and the flexoelectric effect, which are both derived from Lifshitz invariants in the framework of the Landau theory of second-order phase transitions [22].

## II. CONSIDERED PROBLEM AND MATERIAL PARAMETERS

Using a LGD phenomenological approach along with electrostatic equations and elasticity theory, we model the polarization, the internal electric field, and the elastic stresses and strains in a core-shell nanoparticle using FEM, where the ferroelectric core is made of $BaTiO_3$ and has a cylindrical shape. The aspect ratio of the nanocylinder radius $R$ to its length $h$ is significantly higher than unity. The z-axis is parallel to the cylinder axis (**Fig. 1**). The shell is an elastically soft paraelectric or high-k semiconductor with a thickness $\Delta R \ll R$ and screening length $\Lambda \geq 1$ nm. The coverage can be artificial (e.g., a soft organic semiconductor or vacancy-enriched $SrTiO_3$) or natural, and in the latter case it would originate from the polarization screening by surrounding media. The core-shell nanoparticle is placed in a very soft elastic medium.

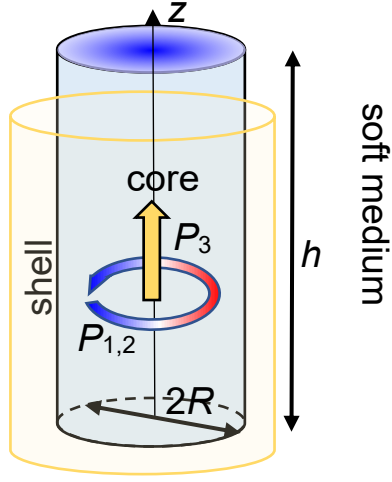

**FIGURE 1.** A cylindrical ferroelectric nanoparticle (core) of radius $R$, covered with an elastically soft semiconducting shell with a thickness $\Delta R \ll R$ and screening length $\Lambda$ of 1 nm, placed in an isotropic elastically soft effective medium. The direction of axial polarization $P_3$ is shown by the straight orange arrow, and lateral components $P_{1,2}$ are shown by the curled red-blue arrow to highlight their vortex-type structure.

The LGD free energy functional $G$ of the nanoparticle core includes a Landau energy – an expansion on powers of 2-4-6 of the polarization ($P_i$), $G_{Landau}$; a polarization gradient energy, $G_{grad}$; an electrostatic energy, $G_{el}$; an elastic, electrostriction contribution $G_{es}$, a flexoelectric contribution, $G_{flexo}$; and a surface energy, $G_S$. It has the form [51]:

$$G = G_{Landau} + G_{grad} + G_{el} + G_{flexo} + G_{flexo} + G_S, \qquad (1a)$$

$$G_{Landau} = \int_{V_C} d^3r \left[ a_i P_i^2 + a_{ij} P_i^2 P_j^2 + a_{ijk} P_i^2 P_j^2 P_k^2 \right], \qquad (1b)$$

$$G_{grad} = \int_{V_C} d^3r \frac{g_{ijkl}}{2} \frac{\partial P_i}{\partial x_j} \frac{\partial P_k}{\partial x_l}, \qquad (1c)$$

$$G_{el} = -\int_{V_C} d^3r \left( P_i E_i + \frac{\varepsilon_0 \varepsilon_b}{2} E_i E_i \right), \qquad (1d)$$

$$G_{es} = -\int_{V_C} d^3r \left( \frac{s_{ijkl}}{2} \sigma_{ij} \sigma_{kl} + Q_{ijkl} \sigma_{ij} P_k P_l \right), \qquad (1e)$$

$$G_{flexo} = -\int_{V_C} d^3r \frac{F_{ijkl}}{2} \left( \sigma_{ij} \frac{\partial P_k}{\partial x_l} - P_k \frac{\partial \sigma_{ij}}{\partial x_l} \right), \qquad (1f)$$

$$G_S = \frac{1}{2} \int_S d^2r\, a_{ij}^{(S)} P_i P_j. \qquad (1g)$$

Here $V_C$ is the core volume. The coefficient $a_i$ linearly depends on temperature $T$, $a_i(T) = \alpha_T [T - T_C]$, where $\alpha_T$ is the inverse Curie-Weiss constant and $T_C$ is the ferroelectric Curie temperature renormalized by surface tension/intrinsic surface stresses [52, 53, 54] and surface bond contraction [55, 56]. Tensor components $a_{ij}$ are regarded as temperature-independent. The tensor $a_{ij}$ is positively defined if the ferroelectric material undergoes a second order transition to the paraelectric phase and negative otherwise. The higher nonlinear tensor $a_{ijk}$ and the gradient coefficients tensor $g_{ijkl}$ are positively defined and regarded as temperature-independent. In Eq.(1e), $\sigma_{ij}$ is the stress tensor, $s_{ijkl}$ is the elastic compliances tensor, and $Q_{ijkl}$ is the electrostriction tensor. In the Lifshitz invariant, Eq.(1f), $F_{ijkl}$ is the flexoelectric tensor.

Landau-Khalatnikov equations [57, 58] obtained from a variation of the free energy (1), mathematical formulation of the electrostatic and elastic sub-problem (see e.g. [59]), initial and boundary conditions (see e.g. [60, 61, 62]), sensitivity to the shape of the cylinder ends, polarization gradient coefficients, shell dielectric permittivity and semiconducting properties, and other details of FEM are given in **Appendix A** of **Suppl. Mat.** [63]. The ferroelectric, dielectric, and elastic properties of the BaTiO$_3$ core are collected from Refs. [64, 65, 66, 67, 68 and 69] and given in **Table SI**.

### III. RESULTS OF FINITE ELEMENT MODELING
#### A. FEM Results at Room Temperature

Images in **Figs. 2a** and **3a** are calculated without electrostriction ($Q_{ij} = 0$) and flexoelectric ($F_{ij} = 0$) couplings between the electric polarization and elastic stresses. For the case a very prolate dipolar kernel oriented along z-axis appears inside the cylindrical core. The kernel has relatively thin 180-degree domain walls, which are mostly uncharged because they are parallel to the kernel axis and cylinder lateral surface. The bound charges appear at the walls only in a small spatial region near the kernel that is contact with the cylinder ends, where the 180-degree walls become counter head-to-head walls. The axial polarization $P_3$ inside the kernel is high, $P_3 \sim -(20-25)$ μC/cm$^2$ (this is very close to the bulk polarization of BaTiO$_3$ $\sim 26$ μC/cm$^2$), and the surrounding core has relatively small axial polarization of the opposite sign, $P_3 \sim (0-5)$ μC/cm$^2$. The lateral components of polarization, $P_1$ and $P_2$, form a two-dimensional (2D) vortex without a central empty core, because a dipolar kernel evolves instead (**Fig. S4** [63]). The two symmetrical Bloch points with $\boldsymbol{P} = 0$ are located at the junction of the dipolar kernel with the cylinder ends. The "up" or "down" orientation of polarization component $P_3$ inside the kernel is determined by random noise in the initial conditions.

Images in **Figs. 2b-d** and **3b-e** are calculated for a nonzero electrostriction coupling ($Q_{ij} \neq 0$) and either negative, zero, or positive values of the flexoelectric coefficients $F_{ij}$. In the presence of electrostriction coupling the dipolar kernel disappears completely (**Figs. 2c** and **3c**). The flexoelectric effect induces an axial component of polarization consisting of two oppositely oriented diffuse $P_3$-domains located near the cylinder ends and separated by a region with $P_3 \approx 0$ (**Figs. 2b, 2d** and **3b, 3d**).

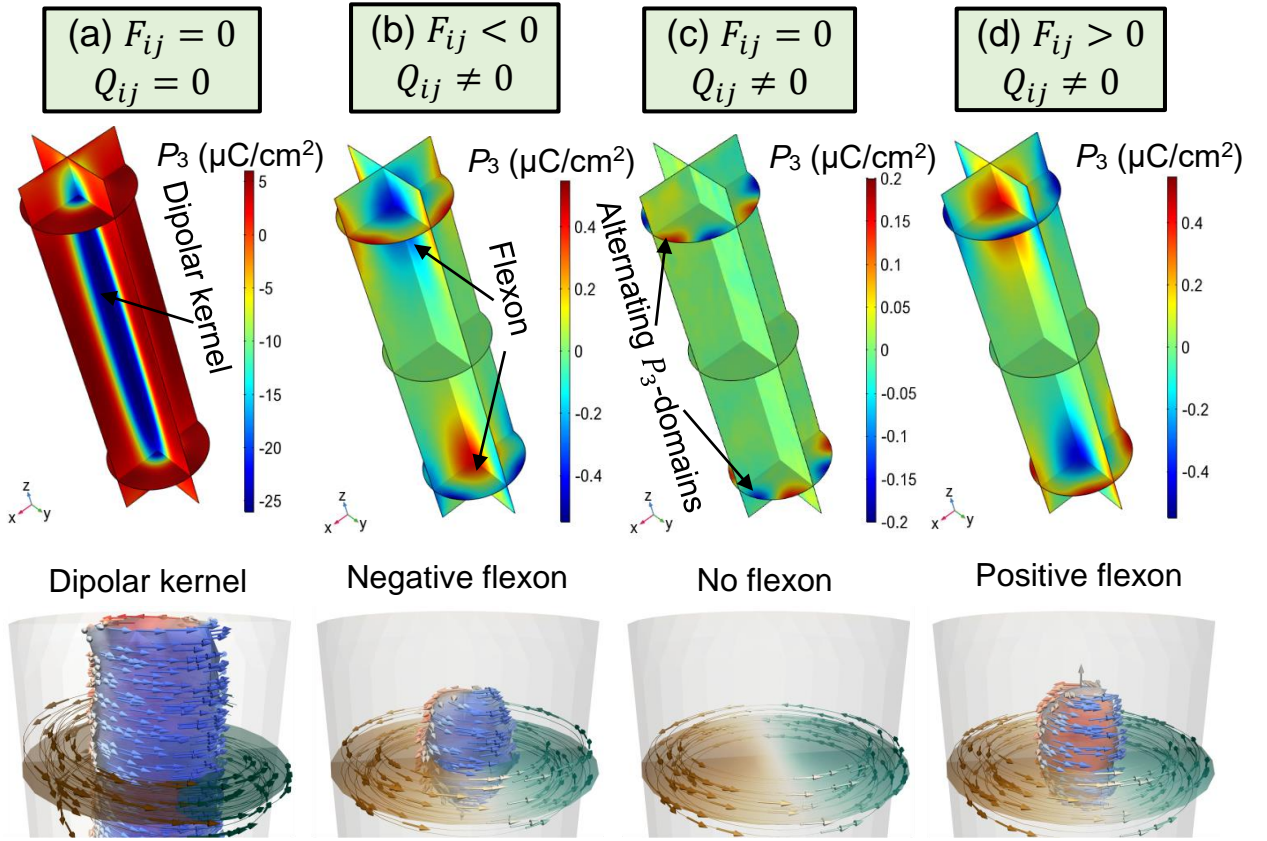

**FIGURE 2**. Distribution of the polarization component $P_3$ (the **top row**) inside a cylindrical nanoparticle and a magnified view on the flexon structure (the **bottom row**). The arrows show the orientation of polarization vector **P**. The images are calculated without electrostriction ($Q_{ij} = 0$) and flexoelectric ($F_{ij} = 0$) couplings (**a**); with electrostriction coupling ($Q_{ij} \neq 0$) and negative (**b**), or zero (**c**), or positive (**d**) values of flexoelectric coefficients $F_{ij}$. The values of $F_{ij}$ and all other parameters are given in **Table SI**, $T = 300$ K. Note the different scales for $P_3$-distributions in plots (a)-(d) in order to maintain a contrast between the different regions.

The diffuseness of the $P_3$-domain walls is dictated by the need to decrease the depolarization field produced by the bound charges of the head-to-head domain walls. The $P_3$-domains are located near the cylinder ends, and their length (about 10 nm) and lateral size (about 5 nm) are almost independent on the cylinder length if $h \gg 5$ nm. The component $P_3$ is very small ($|P_3| \leq 0.4$ μC/cm$^2$), but it increases up to 1.2 μC/cm$^2$ with the flexoelectric coupling increase (**Figs. 3e**) and then saturates (**Figs. 3f**). The axial $P_3$-domains, which have opposite direction of polarization, change their direction under the transformation $F_{ij} \to -F_{ij}$ (compare the position of red and blue diffuse spots of the $P_3$ distributions in **Fig. 3b** and **3d**), while the distribution of the lateral components $P_{1,2}$ and the polarization magnitude $P$ are virtually independent of the $F_{ij}$ sign and magnitude (**Figs. S4-5** [63]).

The maximal ($P_{max}$) and minimal ($P_{min}$) values of $P_3$ are shown by the red and blue curves in **Fig. 3f**. The values $P_{max}$ and $P_{min}$ are even functions of the flexoelectric coupling strength $f$, where $F_{ij} = fF_{ij}^0$ and the reference values of $F_{ij}^0$ are given in **Table SI**. The extremal (maximal or minimal) value $P_e$ in the center of the diffuse axial $P_3$-domain is an odd function of $f$, which is zero at $F_{ij} = 0$ (the green curve in **Fig. 3f**). Note that the $P_e$ value frequently differs from $P_{max}$ and $P_{min}$ values due to the presence of the small sixteen $P_3$-domains localized near the top and bottom junction of the sidewall with the cylinder ends (bottom row in **Fig. S5** [63]).

For the remainder of the paper, we refer to the localized polarization structure near the wire ends as a "**flexon**" for the sake of brevity and to underline the flexoelectric nature of its axial polarization. The main effect of a change of sign in the flexoelectric coefficients is the reorientation of the flexon axial polarization. The polarization structures at the wire ends shown in **Fig. 2b**-**2d** and **Fig. 3b**-**2d** display localized chiral structures with different chirality on opposite ends of the wire, and their chirality changes upon reversal of the sign of the flexoelectric coupling constant.

To understand the chirality change, we derived in **Appendix E** [63] an approximate analytical expression for the polarization distribution inside the flexon:

$$P_1(\rho,\varphi,z) \approx p(\rho,z)\sin\varphi, \quad P_2(\rho,\varphi,z) \approx -p(\rho,z)\cos\varphi, \tag{2a}$$

$$P_3(\rho,\varphi,z) \approx \frac{\frac{Q_{44}}{s_{44}}p(\rho,z)[u_{13}(\rho,\varphi,z)\sin\varphi - u_{23}(\rho,\varphi,z)\cos\varphi] - \frac{F_{11}-F_{44}-F_{12}}{s_{11}-s_{12}}\frac{\partial}{\partial z}u_{33}(\rho,\varphi,z)}{2\left[a_1 - \frac{Q_{11}+2Q_{12}}{s_{11}+2s_{12}}p^2(\rho,z) - \frac{Q_{11}-Q_{12}}{s_{11}-s_{12}}u_{33} + \left[g_{11} + (F_{11}-F_{44}-F_{12})\frac{F_{11}-F_{12}}{s_{11}-s_{12}}\right]L_C^z + g_{44}L_C^x\right]}, \tag{2b}$$

where $\{\rho,\varphi,z\}$ are cylindrical coordinates, the function $p(\rho,z) \sim \tanh\left(\frac{\rho}{L_C^x}\right)$, $L_C^x$ and $L_C^z$ are lateral and axial correlation lengths. The functions $u_{ij}(\rho,\varphi,z)$ are elastic strains, $s_{ij}$ are elastic compliances; $Q_{ij}$ are electrostriction tensor components, $g_{ij}$ are polarization gradient coefficients written in Voigt notations. The first term in Eq.(2b) is induced by the electrostriction coupling, and the second term, proportional to $\frac{F_{11}-F_{44}-F_{12}}{s_{11}-s_{12}}\frac{\partial}{\partial z}u_{33}(\rho,\varphi,z)$, is the flexon.

In order to quantify the chirality of the polarization structure and its variation along the cylinder axis, in **Appendix F** [63] we calculate the topological index $n = \frac{1}{4\pi}\int_S \vec{p}\left[\frac{\partial \vec{p}}{\partial x} \times \frac{\partial \vec{p}}{\partial y}\right]dxdy$ [70] of the unit polarization orientation $\vec{p} = \frac{\vec{P}}{P}$ for the integration over the cylinder cross-section $\{x, y\}$. For the case of $P_3(\rho = R, z) \to 0$, z-dependence of the topological index is

$$n(z) = -\frac{P_3(\rho=0,z)}{2P(\rho=0,z)} \cong -\frac{\text{sign}[f]z}{2\sqrt{1+(z^2/B)}}. \qquad (3)$$

Here $\text{sign}[f]$ is the sign of the flexoelectric coefficients $F_{ij}$, $B$ is a positive constant, which depends on the absolute value of $|F_{ij}|$. $n(z)$ is a normalized profile of $P_3(\rho = 0, z)$, and so $n(z) = 0$ for $F_{ij} = 0$, and its sign is defined by the sign of $F_{ij}$. The dependence $n(z)$ is shown in **Fig. 3g** and **Fig. S10a** for zero, positive, and negative $F_{ij}$. Since the value $P(0, z)$ is very close to the $P_3(0, z)$ near the cylinder ends (**Fig. S10a** and **S10b**), and $P_3(0, z)$ vanishes in the central part of a nanoparticle, the topological index continuously changes from -½ to +½ with a z-coordinate change from one cylinder end to the other. The result clearly shows the localization of the chiral structures – the flexons – at the ends of the wires. The topological index, which can be interpreted as the degree to which a structure is chiral, changes sign from one end to the other, and changes sign upon reversal of the sign of $F_{ij}$. It also increases in magnitude with increasing absolute value of $|F_{ij}|$. These properties are evidence of an obvious correlation between the flexoelectric effect and the formation of chiral polarization structures.

The revealed type of isolated chiral polarization structures, i.e., flexons, display topological features of a three-dimensional meron. In this sense, the polarization vortex in the XY-plane can be interpreted as the Bloch-like transition region of a meron connecting polarization directions of opposite $P_3$ sign in the core region and in the outer cylindrical shell (**Fig. 2**). The flexon polarization $\vec{P}$ develops a characteristic drop-shape with a **chiral** structure localized near the surfaces of the cylinder that is reminiscent of the chiral-bobber state found in non-centrosymmetric magnetic films [71] and nanoparticles [72]. It is worth noting that similar, skyrmion-like configurations at the ends of cylindrical nanowires have also been predicted analytically [73] and numerically [74] in the case of non-chiral ferromagnetic materials, but only in the form of transient configurations appearing during the dynamic magnetization reversal process. Here, the skyrmion-like polarization structures appear as stable states in the ferroelectrics, owing to a chiral-symmetry breaking effect of the flexoelectric coupling. In contrast to previous findings [75, 76, 77], the flexon structure is chiral [78] and almost uncharged because $div\vec{P} \cong 0$ (**Fig. S9b** [63]).

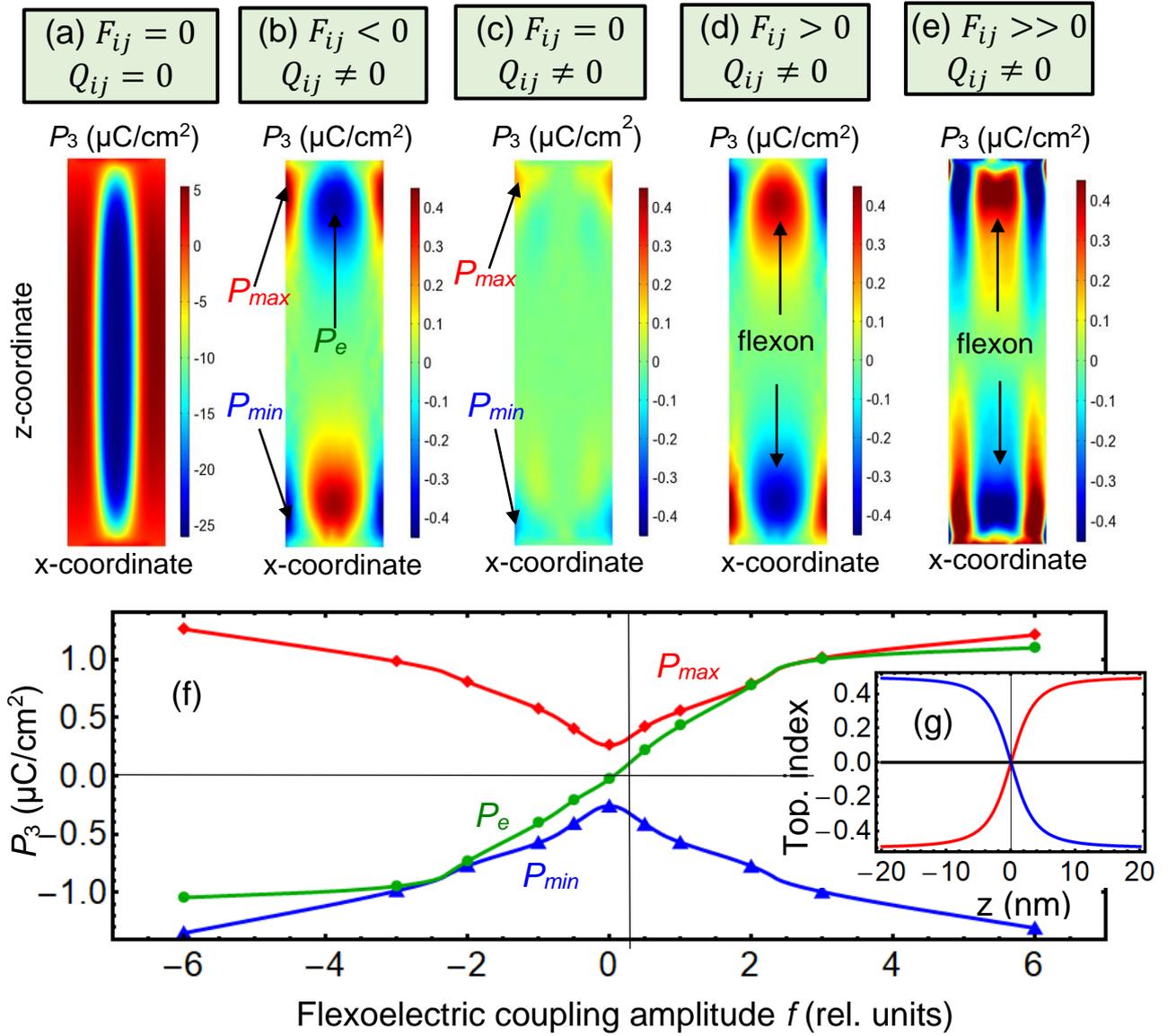

**FIGURE 3**. Distribution of the polarization component $P_3$ (the **top row**) in the XZ cross-section of the nanoparticle. Images are calculated without electrostriction ($Q_{ij} = 0$) and flexoelectric ($F_{ij} = 0$) coupling **(a)**; with electrostriction coupling ($Q_{ij} \neq 0$) and negative **(b)**, zero **(c)**, positive **(d)**, or high positive **(e)** values of flexoelectric coefficients $F_{ij}$. The **bottom part (f)** is the dependence of the maximal (red curve, $P_{max}$) and minimal (blue curve, $P_{min}$) values of $P_3$ on the relative amplitude of the flexoelectric coupling strength $f$. The green curve is the extremal (maximal or minimal) value $P_e$ in the center of the top axial $P_3$-domain. Here $F_{ij} = f F_{ij}^0$, the values of $F_{ij}^0$ and all other parameters are given in **Table SI**, $T = 300$ K. The Z-profile of the polarization topological index $n(z)$ is shown in the inset **(g)** for zero (black line), positive (red curve), and negative (blue curve) $F_{ij}$. Note the different scales for $P_3$ in the plots (a) and (b)-(e) in order to maintain a contrast between the different regions.

As a rule, the flexoelectric tensor component $F_{44}$ is either poorly known from experiments or ill-defined from ab initio calculations; therefore, we can vary it over a wide range to determine the degree by which the flexoelectric coupling anisotropy influences the morphology of the polarization state. Corresponding FEM results are shown in **Fig. 4.** The top and middle rows illustrate that the $P_3$ distribution changes very strongly when $F_{44}$ varies from high negative to high positive values, while the other components of the flexoelectric tensor are fixed and equal to the tabulated values $F_{11} = 2.4 \cdot 10^{-11}$ m$^3$/C and $F_{12} = 0.5 \cdot 10^{-11}$ m$^3$/C.

The flexon contains two pronounced axial domains located near the cylinder ends, which have thick diffuse domain walls and opposite polarization directions, and exist at high negative (**Fig. 4a**) and high positive (**Fig. 4b** and **4e**) $F_{44}$ values. The $P_3$-domains become smaller and more diffuse with a decrease of $|F_{44}|$; but they are still visible and practically do not change their shape, size, or polarization distribution for small $|F_{44}|$ values over the range $|F_{44}| \leq 0.06$ (**Fig. 4b**). The flexon becomes faint and almost disappears when $F_{44}$ approaches the value $F_{44} = F_{11} - F_{12} = 1.9 \cdot 10^{-11}$ m$^3$/C corresponding to the isotropic symmetry of $F_{ij}$ (**Fig. 4c**). The value will be referred to as "isotropic" below.

The dependence of the maximal (red curve, $P_{max}$) and minimal (blue curve, $P_{min}$) values of the polarization component $P_3$ on the relative amplitude $f$ of the flexoelectric coefficient $F_{44}$ is shown in **Fig. 4f**, where $F_{44} = f F_{44}^0$ and $F_{44}^0 = 0.06 \cdot 10^{-11}$ m$^3$/C. The values $P_{max}$ and $P_{min}$ reach a very diffuse plateau-like minimum and maximum, respectively, at the isotropic value $F_{44} = F_{11} - F_{12}$. The green curve in **Fig. 4f** is the extremal value $P_e$ in the center of the bottom axial $P_3$-domain. The extremal value $P_a$ in the center of the diffuse $P_3$-domain changes its sign in the immediate vicinity of the isotropic value $F_{44} = F_{11} - F_{12}$. The values $P_{max}$, $P_{min}$, and $P_e$ have no definite parity, because they are neither odd nor even functions of the flexoelectric coefficient $F_{44}$ amplitude $f$. From **Fig. 4f** we can conclude that the anisotropy of the flexoelectric coupling critically influences the morphology of the flexon, where the axial part of the flexon polarization is proportional to $-\frac{F_{11}-F_{44}-F_{12}}{s_{11}-s_{12}}\frac{\partial u_{33}}{\partial z}$ [Eq.(2b)], this proportionality along with **Fig. S9** qualitatively describes the curves' behavior in **Fig. 4f.**

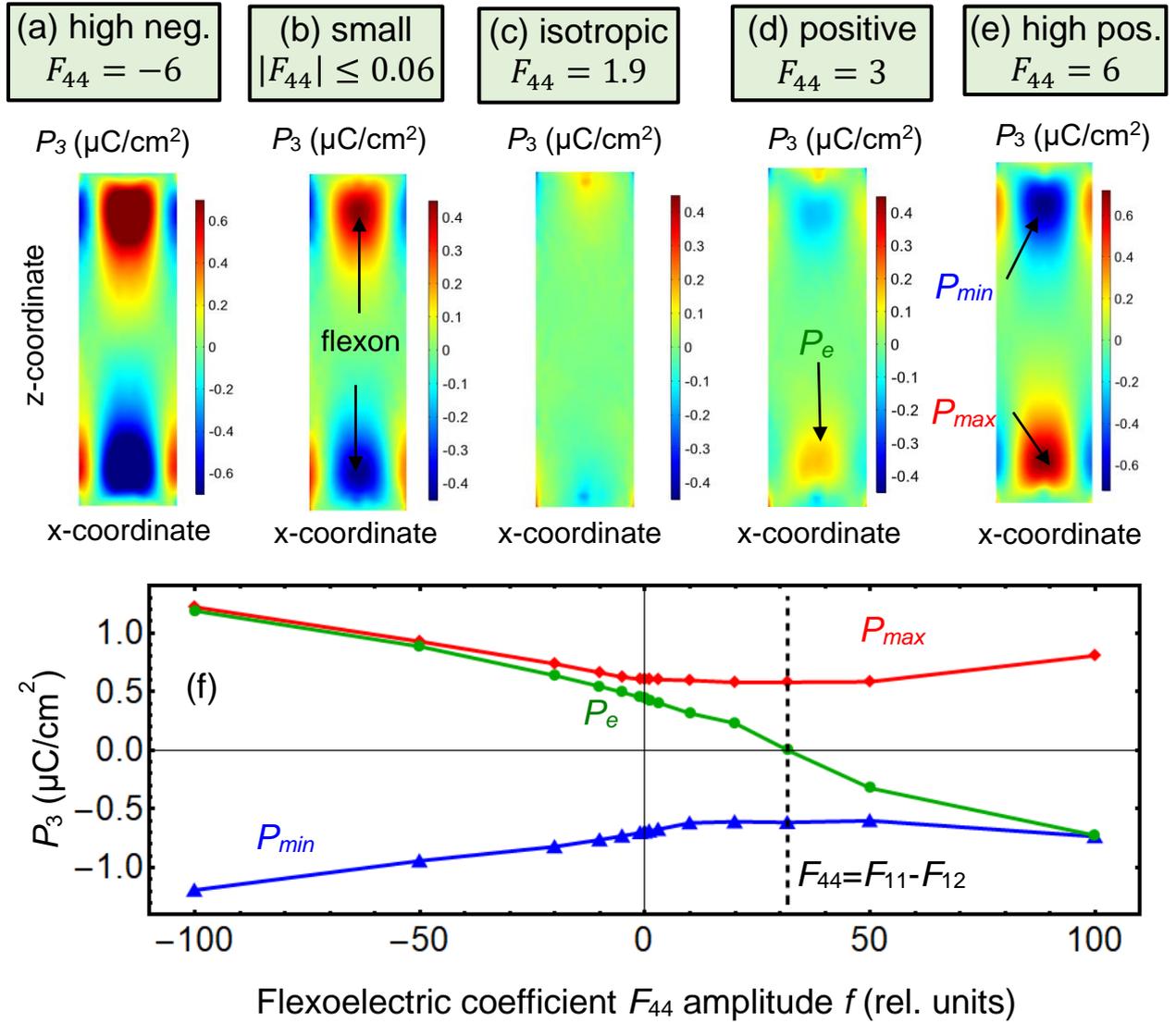

**FIGURE 4**. Distribution of the polarization component $P_3$ in the XZ cross-section of cylindrical core (the **top row**). Images are calculated for the fixed values $F_{11} = 2.4 \cdot 10^{-11}$ m$^3$/C and $F_{12} = 0.5 \cdot 10^{-11}$ m$^3$/C, while the value of $F_{44}$ varies from -6 to 6 (in $10^{-11}$ m$^3$/C) as indicated in the legends. The **bottom part (f)** is the dependence of the maximal (red curve, $P_{max}$) and minimal (blue curve, $P_{min}$) values of $P_3$ on the relative amplitude $f$ of the flexoelectric coefficient $F_{44}$ in the core. The green curve is the extremal (maximal or minimal) value $P_e$ in the center of the bottom axial $P_3$-domain. Here $F_{44} = f F_{44}^0$ and $F_{44}^0 = 0.06 \cdot 10^{-11}$ m$^3$/C. The electrostriction coupling coefficients $Q_{ij}$ and all other parameters are listed in **Table SI**, $T = 300$ K. Note the different scales for $P_3$ in the plots (a)-(e) in order to maintain a contrast between the different regions.

**B. Temperature Behavior of the Flexon-Type Polarization Distribution**

To define the temperature interval in which flexons exist as stable or meta-stable states, we performed FEM in the temperature range from 50 K to 400 K using different initial distribution of polarization in a cylindrical core. Typical FEM results are shown in **Fig. 5**, where the columns (a)-(e) correspond to the temperature increase from 240 K to 370 K; the structure of the azimuthal components of the polarization vector, $P_1$ and $P_2$, is vortex-like and shows weak variations when approaching the surface over the same temperature range (see the direction of arrows at the bottom image of **Figs. S7** [63]).

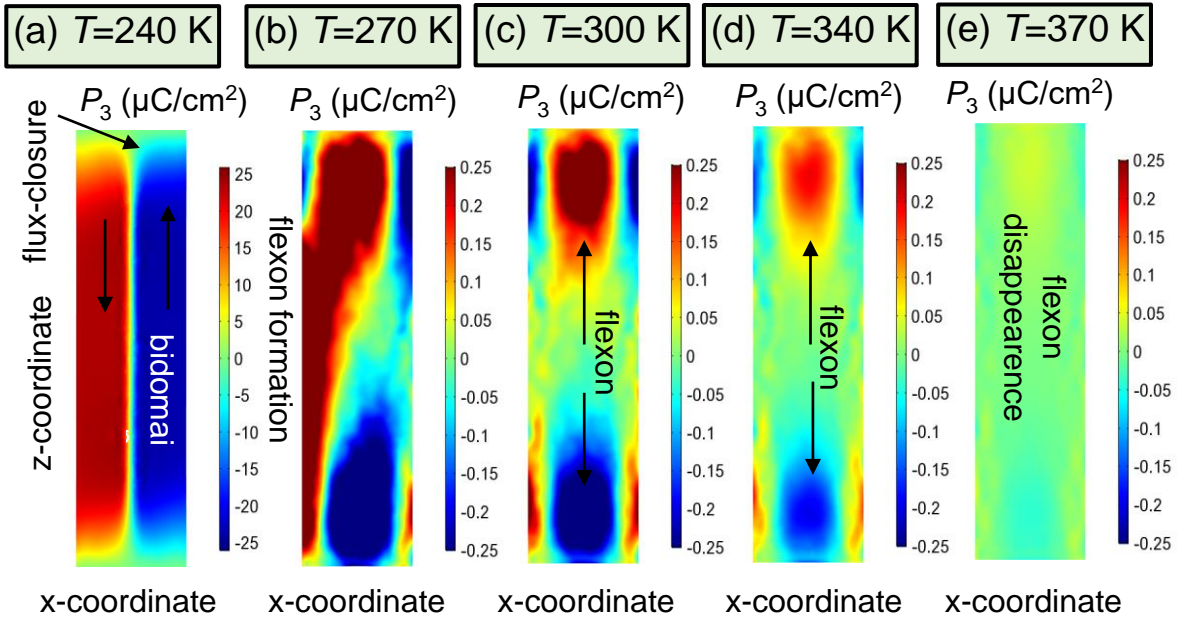

**FIGURE 5**. Distribution of polarization component $P_3$ in XZ cross-sections of the nanoparticle core. Different columns are calculated for the temperatures $T = 240, 270, 300, 340,$ and $370$ K (**a, b, c, d, e**). All other parameters are listed in **Table I**. Note the different scales for $P_3$ in the plots (a) and (b)-(e) in order to maintain a contrast between the different regions.

A bidomain configuration of $P_3$ is stable at temperatures lower than 250 K (**Figs. 5a**). The bidomain structure has a relatively thin uncharged 180-degree domain wall inside the cylinder, which transforms into a flux-closure domain near the electrically-open cylinder ends. An initial four-domain polarization distribution relaxes to a flexon-like domain structure in the temperature range $260\ \text{K} < T < 360\ \text{K}$ (**Figs. 5b-d**). The flexon gradually disappears at $T > 370$ K (the middle image in **Figs. 5e**). The ferroelectric polarization inside the core significantly decreases at $T > 370$ K and completely disappears at $T \sim 400$ K (the top image in **Figs. 5e**). The structure becomes faint with a temperature increase above 370 K (**Figs. 5e**), hence, the flexon-type polarization distribution exists in a relatively wide temperature range $260\ \text{K} < T < 360\ \text{K}$.

The axial counter domains inherent to flexons are the most pronounced feature over the narrower range 290 K $< T <$ 340 K.

## IV. DISCUSSION

Any deviation from a four quadrants domain configuration in the flexon-type polarization distribution is found to be metastable. This is because the antiparallel bidomain-type polarization distribution (starting from a random noise) has a lower free energy in a BaTiO$_3$ cylindrical nanoparticle. The derived energy values at room temperature are $G_{fl} = -3.6 \cdot 10^{-18}$ J and $G_{bd} = -4.0 \cdot 10^{-18}$ J in the flexon and the bidomain structure, respectively. The energy difference between these states, $\Delta G = 4 \cdot 10^{-19}$ J is much higher (about $100\, k_B T$) than the thermal energy barrier $k_B T$ at room temperature. However, the ratio $\frac{\Delta G}{k_B T}$ strongly decreases with as the temperature increases. The linear relative dielectric permittivity in both states is about 110 at room temperature and it strongly increases with temperature. Furthermore, our numerical simulations show that a spontaneous off-field transition from the flexon to the bidomain polarization state does not occur, whereas the in-field transition is possible (corresponding hysteresis loops are shown in **Appendix D** [63]). Thus, the bidomain and flexon states of a prolate core-shell ferroelectric nanoparticle can be considered as the exited and ground state of a two-level system suitable for information recording. The two-level system can imitate qubits operating in the temperature range where $1 < \frac{\Delta G}{k_B T} < 5$. Furthermore, the bidomain polarization state corresponds to an antiferroelectric-like state of the nanoparticle polarization, which can be represented as two antiparallel nanoscopic dipoles. The flexon is a much more complex achiral vortex-like configuration containing two counter dipole nanodomains with diffuse relaxor-like polar properties. Thus, an ensemble of prolate core-shell ferroelectric nanoparticles, where a given nanoparticle is either in a flexon or a bidomain state, can be an alternative media for information processing. The media may exhibit unusual properties including antiferroelectric and/or relaxor-like polarization states, which can lead to additional functionalities. Note that the appearance of the antiferroelectric and relaxor-like glass states, as well as a newly discovered liquid glass state [79] with additional (anti)ferroelectric ordering and other cross-talk effects, are possible in a suspension of the prolate core-shell ferroelectric nanoparticles.

The relatively wide temperature range (about 100 K) corresponding to the stability or meta-stability of the flexon-type polarization distribution gives us the hope that the domain morphology can be observed experimentally. Specifically, the measurements of local vertical displacement by piezoresponse force microscopy (**PFM**) visualize the distribution of $P_3(\vec{r})$ at distances ~ 10 nm from the ends of a nanoparticle, but the resolution procedure for the local piezoresponse of diffuse domains under the surface is far not straightforward [80, 81]. This is because PFM is a near-field method. A complementary tool to probe chiral polar textures is far-field nonlinear optical microscopy [82], which has a comparatively much lower resolution than PFM, yet optimum focusing methods and the experimental geometry allow for overcoming the diffraction limit. For example, second-harmonic generation microscopy was successfully used by the community to precisely study semiconductor nanowires [83]. This method should also be capable of providing complementary information on the 3D ferroelectric domain structure (being sensitive to $P_{1,2}$ and $P_3$) by using polarimetry analysis (see, e.g., [84]). Another promising method is resonant elastic soft X-ray scattering, a synchrotron-based method sensitive to chiral polar arrangements through dichroism effects [85, 86]. This method was successfully applied to detect different topological structures, including vortices [36], skyrmions [41], and chiral domain walls [87].

Note that Liu et al. [39] revealed that an axial polarization component of the vortex can appear in ferroelectric $PbTiO_3$ nanodots due to the flexoelectric effect. Thus, the paper [39] and this work predict different flexo-sensitive vortex-like states with an axial polarization in ferroelectrics nanoparticles of various geometry. At that this work reveals the critical influence of the flexoelectric coefficients sign and anisotropy on the appearance and properties of the axial polarization, and, most important, on the chirality of a ferroelectric vortex. Qualitatively, both works, [39] and this one, illustrate that flexoelectricity can change the chiral state of a polarization texture, and this work studies the behavior of a topological index (in fact a skyrmion number) that quantifies the change (see **Fig. 3g**).

The main features characterizing polarization structures stabilized by DMI-type interactions are their breaking of chiral symmetry and their incommensurability, i.e., a long-period modulation in space that is unrelated to the crystalline lattice parameter. The appearance of such chiral incommensurate phases can generally be attributed to specific energy terms in the Landau-Ginzburg potential, known as Lifshitz invariants [49, 88]. Only a few ferroelectrics have crystalline structures whose symmetry allows such Lifshitz invariants; therefore, an interaction directly analogous to the magnetic DMI is generally not found in ferroelectrics. However, although not identical in its mathematical form, the energy density due to the flexoelectric coupling [23] is similar to a DMI-type energy term related to a linear Lifshitz invariant. Accordingly, we argue that the flexoelectric coupling can lead to polarization states with properties similar to those generated by a DMI-type interaction. Such a connection between flexoelectric coupling, Lifshitz invariants, and DMI has been discussed before in the case of liquid crystals [89]. It was found that flexoelectricity in liquid crystals can play a central role in the development of modulated phases that are analogous to those known from chiral ferromagnets [90]. Our study shows that –similar to liquid crystals, where elastic strains fields couple to electric fields– the flexoelectric effect through which strain fields couple to the electric polarization field can lead to comparable modulated phases with chiral properties in a ferroelectric nanoparticle.

## V. SUMMARY AND CONCLUSIONS

Using FEM simulations based on the phenomenological LGD approach alongside electrostatic equations and elasticity theory, we identify a characteristic polarization structure developing between two oppositely oriented axial domains located near the cylinder ends. This polar structure, which we named "flexon", displays chiral features that are connected to the flexoelectric coupling. In the azimuthal plane, the flexon polarization forms a localized chiral structure resembling a meron, or a vortex with a central kernel. Analytical calculations and FEM prove that the flexon axial polarization, and thus its chirality, switches upon a change of the sign of the flexoelectric coefficients. We also observe that the anisotropy of the flexoelectric coupling critically influences the flexon formation and the related domain morphology. This observation corroborates the link between chirality and flexocoupling, and it identifies the flexoelectric effect as the driving force stabilizing these structures.

While in magnetic systems with strong DMI, similar localized chiral structures have been reported [71], the polarization state discussed here is formed without any ferroelectric counterpart of the DMI [46]. We recall that, like the DMI, the flexoelectric coupling is derived as a Lifshitz invariant [Eq. (1f)] in the context of the Landau theory of phase transitions [22], and that such linear Lifshitz invariants generally play a key role in the formation of helical structures [49, 88]. The fact that both the DMI and the flexoelectric stabilize structures with a specific chirality demonstrates an analogy between these two interactions which appears to have been overlooked in the literature of ferroelectric solids. An important difference compared to the classical DMI is that the flexoelectric coupling is ubiquitous in ferroelectrics, whereas the ferroelectric DMI is forbidden by symmetry in most material types. Therefore, a chiral interaction mediated by the flexoelectric effect can potentially be found in all ferroelectrics. The coupling of the electric polarization and elastic strain gradients could thus be a much more commonly accessible alternative interaction for the formation of chiral and achiral structures [75, 76]. This coupling could also open the possibility of generating and dissolving chiral polarization states through strain engineering [91].

We predict that the pronounced flexon-type polarization distribution with two axial counter domains exists in the temperature range $290 \text{ K} < T < 340 \text{ K}$. The relatively wide temperature range (about 50 K) corresponding to the stability or meta-stability of the flexon-type polarization distribution give us the hope that the flexons can be observed experimentally. However, the analysis of the hysteresis loops leads to the conclusion that flexons and other domain configurations cannot be resolved from macroscopic measurements of the average polarization in a homogeneous electric field. We anticipate that flexons can be reliably observed, e.g., by the local methods using a strong gradient of electric field, such as PFM, which gives us the information about the distribution of polarization with a nanoscale resolution.

**Acknowledgements.** A.N.M. acknowledges EOARD project 9IOE063 and related STCU partner project P751. R.H. and S.C.-H. acknowledge funding from the French National Research Agency through contract ANR-18-CE92-0052 "TOPELEC". V.Y.R. acknowledges the support of COST Action CA17139. A portion of FEM was conducted at the Center for Nanophase Materials Sciences, which is a DOE Office of Science User Facility (CNMS Proposal ID: CNMS2021-B-00843).

## Supplementary Materials to

## "Chiral Polarization Textures Induced by the Flexoelectric Effect in Ferroelectric Nanocylinders"

### APPENDIX A. Mathematical Formulation of the Problem and FEM Details
### A. Mathematical Formulation of the Problem

We use the Landau-Ginzburg-Devonshire (LGD) approach combined with electrostatic equations, because this method has proven to be successful in establishing the physical origin of anomalies in phase diagrams, determining polar and dielectric properties of ferroelectric nanoparticles [1, 2], and calculating the changes of their domain structure morphology with size reduction [3, 4]. The LGD approach allows for the consideration of various size and surface effects, such as correlation effects and depolarization fields arising in the case of incomplete polarization screening [5], surface bond contraction [6, 7], and intrinsic surface stresses and strains [8, 9, 10].

The LGD free energy functional $G$ additively includes a Landau expansion on powers of 2-4-6 of the polarization ($P_i$), $G_{Landau}$; a polarization gradient energy contribution, $G_{grad}$; an electrostatic contribution, $G_{el}$; the elastic, electrostriction, flexoelectric contributions, $G_{es+flexo}$; and a surface energy, $G_S$. It has the form [3, 10, 11]:

$$G = G_{Landau} + G_{grad} + G_{el} + G_{es+flexo} + G_S, \qquad (A.1a)$$

$$G_{Landau} = \int_{V_c} d^3r \left[ a_i P_i^2 + a_{ij} P_i^2 P_j^2 + a_{ijk} P_i^2 P_j^2 P_k^2 \right], \qquad (A.1b)$$

$$G_{grad} = \int_{V_c} d^3r \frac{g_{ijkl}}{2} \frac{\partial P_i}{\partial x_j} \frac{\partial P_k}{\partial x_l}, \qquad (A.1c)$$

$$G_{el} = -\int_{V_c} d^3r \left( P_i E_i + \frac{\varepsilon_0 \varepsilon_b}{2} E_i E_i \right) - \frac{\varepsilon_0}{2} \int_{V_s} \varepsilon_{ij}^S E_i E_j d^3r - \frac{\varepsilon_0}{2} \int_{V_o} \varepsilon_{ij}^e E_i E_j d^3r, \qquad (A.1d)$$

$$G_{es+flexo} = -\int_{V_c} d^3r \left( \frac{s_{ijkl}}{2} \sigma_{ij} \sigma_{kl} + Q_{ijkl} \sigma_{ij} P_k P_l + F_{ijkl} \left( \sigma_{ij} \frac{\partial P_k}{\partial x_l} - P_k \frac{\partial \sigma_{ij}}{\partial x_l} \right) \right) \qquad (A.1e)$$

$$G_S = \frac{1}{2} \int_S d^2r \, a_{ij}^{(S)} P_i P_j. \qquad (A.1f)$$

Here $V_c$ and $V_s$ are the core and shell volume, respectively. The coefficient $a_i$ linearly depends on temperature $T$, $a_i(T) = \alpha_T [T - T_C]$, where $\alpha_T$ is the inverse Curie-Weiss constant and $T_C$ is the ferroelectric Curie temperature renormalized by electrostriction and surface tension. Tensor components $a_{ij}$ are regarded as temperature-independent. The tensor $a_{ij}$ is positively defined if the ferroelectric material undergoes a second order transition to the paraelectric phase and negative otherwise. The higher nonlinear tensor $a_{ijk}$ and the gradient coefficients tensor $g_{ijkl}$ are positively defined and regarded as temperature-independent. The following designations are used in Eq.(A.1e):

$\sigma_{ij}$ is the stress tensor, $s_{ijkl}$ is the elastic compliances tensor, $Q_{ijkl}$ is the electrostriction tensor, and $F_{ijkl}$ is the flexoelectric tensor.

For cubic (m3m) point symmetry group of the parent phase the explicit form of the "half" Lifshitz invariant for the flexoeffect is

$$\Delta G_{flexo} = [\sigma_{11}F_{11} + (\sigma_{22} + \sigma_{33})F_{12}]\frac{\partial P_1}{\partial x_1} + [\sigma_{22}F_{11} + (\sigma_{11} + \sigma_{33})F_{12}]\frac{\partial P_2}{\partial x_2} + [\sigma_{33}F_{11} + (\sigma_{11} + \sigma_{22})F_{12}]\frac{\partial P_3}{\partial x_3} + F_{44}\left[\sigma_{12}\left(\frac{\partial P_1}{\partial x_2} + \frac{\partial P_2}{\partial x_1}\right) + \sigma_{13}\left(\frac{\partial P_1}{\partial x_3} + \frac{\partial P_3}{\partial x_1}\right) + \sigma_{23}\left(\frac{\partial P_2}{\partial x_3} + \frac{\partial P_3}{\partial x_2}\right)\right] \quad (A.2)$$

Allowing for the Khalatnikov mechanism of polarization relaxation [12], minimization of the free energy (A.1) with respect to polarization leads to three coupled time-dependent Euler-Lagrange equations for polarization components inside the core, $\frac{\delta G}{\delta P_i} = -\Gamma\frac{\partial P_i}{\partial t}$, where $i = 1, 2, 3$. The explicit form of the equations for a ferroelectric crystal with m3m parent symmetry is:

$$\Gamma\frac{\partial P_1}{\partial t} + 2P_1(a_1 - Q_{12}(\sigma_{22} + \sigma_{33}) - Q_{11}\sigma_{11}) - Q_{44}(\sigma_{12}P_2 + \sigma_{13}P_3) + 4a_{11}P_1^3 + 2a_{12}P_1(P_2^2 + P_3^2) + 6a_{111}P_1^5 + 2a_{112}P_1(P_2^4 + 2P_1^2P_2^2 + P_3^4 + 2P_1^2P_3^2) + 2a_{123}P_1P_2^2P_3^2 - g_{11}\frac{\partial^2 P_1}{\partial x_1^2} - g_{44}\left(\frac{\partial^2 P_1}{\partial x_2^2} + \frac{\partial^2 P_1}{\partial x_3^2}\right) = -F_{11}\frac{\partial \sigma_{11}}{\partial x_1} - F_{12}\left(\frac{\partial \sigma_{22}}{\partial x_1} + \frac{\partial \sigma_{33}}{\partial x_1}\right) - F_{44}\left(\frac{\partial \sigma_{12}}{\partial x_2} + \frac{\partial \sigma_{13}}{\partial x_3}\right) + E_1$$

(A.3a)

$$\Gamma\frac{\partial P_2}{\partial t} + 2P_2(a_1 - Q_{12}(\sigma_{11} + \sigma_{33}) - Q_{11}\sigma_{22}) - Q_{44}(\sigma_{12}P_1 + \sigma_{23}P_3) + 4a_{11}P_2^3 + 2a_{12}P_2(P_1^2 + P_3^2) + 6a_{111}P_2^5 + 2a_{112}P_2(P_1^4 + 2P_2^2P_1^2 + P_3^4 + 2P_2^2P_3^2) + 2a_{123}P_2P_1^2P_3^2 - g_{11}\frac{\partial^2 P_2}{\partial x_2^2} - g_{44}\left(\frac{\partial^2 P_2}{\partial x_1^2} + \frac{\partial^2 P_2}{\partial x_3^2}\right) = -F_{11}\frac{\partial \sigma_{22}}{\partial x_2} - F_{12}\left(\frac{\partial \sigma_{11}}{\partial x_2} + \frac{\partial \sigma_{33}}{\partial x_2}\right) - F_{44}\left(\frac{\partial \sigma_{12}}{\partial x_1} + \frac{\partial \sigma_{23}}{\partial x_3}\right) + E_2$$

(A.3b)

$$\Gamma\frac{\partial P_3}{\partial t} + 2P_3(a_1 - Q_{12}(\sigma_{11} + \sigma_{22}) - Q_{11}\sigma_{33}) - Q_{44}(\sigma_{13}P_1 + \sigma_{23}P_2) + 4a_{11}P_3^3 + 2a_{12}P_3(P_1^2 + P_2^2) + 6a_{111}P_3^5 + 2a_{112}P_3(P_1^4 + 2P_3^2P_1^2 + P_2^4 + 2P_2^2P_3^2) + 2a_{123}P_3P_1^2P_2^2 - g_{11}\frac{\partial^2 P_3}{\partial x_3^2} - g_{44}\left(\frac{\partial^2 P_3}{\partial x_1^2} + \frac{\partial^2 P_3}{\partial x_2^2}\right) = -F_{11}\frac{\partial \sigma_{33}}{\partial x_3} - F_{12}\left(\frac{\partial \sigma_{11}}{\partial x_3} + \frac{\partial \sigma_{22}}{\partial x_3}\right) - F_{44}\left(\frac{\partial \sigma_{13}}{\partial x_1} + \frac{\partial \sigma_{23}}{\partial x_2}\right) + E_3$$

(A.3c)

The temperature-dependent Khalatnikov coefficient $\Gamma$ [13] determines the relaxation time of the polarization $\tau_K = \Gamma/|\alpha|$. Consequently, $\tau_K$ typically varies in the range ($10^{-9} - 10^{-6}$) seconds for temperatures far from $T_C$. As argued by Hlinka et al. [14], we assumed that $g'_{44} = -g_{12}$ in Eqs.(A.3).

The boundary condition for polarization at the core-shell interface $r = R$ accounts for the flexoelectric effect:

$$a_{ij}^{(S)} P_j + \left(g_{ijkl} \frac{\partial P_k}{\partial x_l} - F_{klij}\sigma_{kl}\right) n_j \bigg|_{r=R} = 0 \tag{A.4}$$

where **n** is the outer normal to the surface, $i = 1, 2, 3$. In our FEM studies, we use the so-called "natural" boundary conditions corresponding to $a_{ij}^{(S)} = 0$.

The electric field components $E_i$ in Eq.(A.3) are derived from the electric potential $\varphi$ in a conventional way, $E_i = -\partial \varphi / \partial x_i$. The potential $\varphi_f$ satisfies the Poisson equation in the ferroelectric cylinder (subscript "$f$"):

$$\varepsilon_0 \varepsilon_b \left(\frac{\partial^2}{\partial x_1^2} + \frac{\partial^2}{\partial x_2^2} + \frac{\partial^2}{\partial x_3^2}\right) \varphi_f = \frac{\partial P_i}{\partial x_i}, \quad x_1^2 + x_2^2 \leq R \ \cap \ 0 \leq x_3 \leq h, \tag{A.5a}$$

The electric potential $\varphi_e$ in the external region outside the core-shell nanoparticle satisfies the Laplace equation (subscript "$e$"):

$$\varepsilon_0 \varepsilon_e \left(\frac{\partial^2}{\partial x_1^2} + \frac{\partial^2}{\partial x_2^2} + \frac{\partial^2}{\partial x_3^2}\right) \varphi_e = 0, \quad x_1^2 + x_2^2 > R \ \cup \ x_3 < 0 \ \cup \ x_3 > h, \tag{A.5b}$$

Equations (A.5) are supplemented with the continuity conditions for electric potential and linear screening conditions for the normal components of the electric displacements at the cylinder surface S [4]:

$$(\varphi_e - \varphi_f)\big|_S = 0, \quad \mathbf{n}(\mathbf{D}_e - \mathbf{D}_f)\big|_S = -\frac{\varphi_f}{\Lambda}. \tag{A.6a}$$

The boundary condition (A.6a) corresponds to the surface of the core covered by an ultrathin semiconductor shell with an effective screening length $\Lambda \geq 1$ nm [15, 16, 17]. Note that a screening length greater than 0.1 nm leads to the domain formation in the core. The case $\Lambda \to \infty$ corresponds to an electrical open-circuit condition. We impose an electrical open-circuit condition at the cylinder ends to make the vortex-type polarization energetically favorable. Either charges are absent or the applied voltage is fixed at the boundaries of the computation region:

$$\frac{\partial \varphi_e}{\partial x_l} n_l \bigg|_{x=\pm\frac{L}{2}} = 0, \quad \frac{\partial \varphi_e}{\partial x_l} n_l \bigg|_{y=\pm\frac{L}{2}} = 0, \quad \varphi_e\big|_{z=+\frac{L}{2}} = 0, \quad \varphi_e\big|_{z=-\frac{L}{2}} = V_e. \tag{A.6b}$$

Elastic equations of state follow from the variation of the energy (A.1e) with respect to elastic stress, $\frac{\delta G}{\delta \sigma_{ij}} = -u_{ij}$. In the oversimplified case

$$s_{ijkl}\sigma_{kl} + Q_{ijkl}P_k P_l + F_{ijkl}\frac{\partial P_l}{\partial x_k} = u_{ij}, \quad 0 < r \leq R, \quad 0 < z \leq h, \tag{A.7a}$$

$$\sigma_{ij} = c_{ijkl}u_{kl} - q_{ijkl}P_kP_l - f_{ijkl}\frac{\partial P_l}{\partial x_k}, \quad 0 < r \leq R, \quad 0 < z \leq h, \quad \text{(A.7b)}$$

where $u_{ij}$ is the strain tensor components related to the displacement components $U_i$ in the following way: $u_{ij} = (\partial U_i/\partial x_j + \partial U_j/\partial x_i)/2$. For cubic (m3m) point symmetry group of the parent phase the strain components are:

$$u_{11} = s_{11}\sigma_{11} + s_{12}(\sigma_{22} + \sigma_{33}) + Q_{11}P_1^2 + Q_{12}(P_2^2 + P_3^2) + F_{11}\frac{\partial P_1}{\partial x_1} + F_{12}\left(\frac{\partial P_2}{\partial x_2} + \frac{\partial P_3}{\partial x_3}\right) \quad \text{(A.8a)}$$

$$u_{22} = s_{11}\sigma_{22} + s_{12}(\sigma_{11} + \sigma_{33}) + Q_{11}P_2^2 + Q_{12}(P_1^2 + P_3^2) + F_{11}\frac{\partial P_2}{\partial x_2} + F_{12}\left(\frac{\partial P_1}{\partial x_1} + \frac{\partial P_3}{\partial x_3}\right) \quad \text{(A.8b)}$$

$$u_{33} = s_{11}\sigma_{33} + s_{12}(\sigma_{11} + \sigma_{22}) + Q_{11}P_3^2 + Q_{12}(P_2^2 + P_1^2) + F_{11}\frac{\partial P_3}{\partial x_3} + F_{12}\left(\frac{\partial P_1}{\partial x_1} + \frac{\partial P_2}{\partial x_2}\right) \quad \text{(A.8c)}$$

$$u_{12} = s_{44}\sigma_{12} + Q_{44}P_1P_2 + F_{44}\left(\frac{\partial P_1}{\partial x_2} + \frac{\partial P_2}{\partial x_1}\right) \quad \text{(A.8d)}$$

$$u_{13} = s_{44}\sigma_{13} + Q_{44}P_1P_3 + F_{44}\left(\frac{\partial P_1}{\partial x_3} + \frac{\partial P_3}{\partial x_1}\right) \quad \text{(A.8e)}$$

$$u_{23} = s_{44}\sigma_{23} + Q_{44}P_3P_2 + F_{44}\left(\frac{\partial P_2}{\partial x_3} + \frac{\partial P_3}{\partial x_2}\right) \quad \text{(A.8f)}$$

Equations (A.8) should be considered along with equations of mechanical equilibrium $\partial\sigma_{ij}(\boldsymbol{x})/\partial x_i = 0$ and compatibility equations $e_{ikl}e_{jmn}\partial^2 u_{ln}(\boldsymbol{x})/\partial x_k \partial x_m = 0$, which are equivalent to the mechanical displacement vector $U_i$ continuity [18]. The boundary conditions for elastic stresses are the virtual absence of their normal components at the nanoparticle surface, as the ambient media is regarded absolutely soft.

### B. Finite Element Modelling Details

<u>Electrical boundary conditions.</u> We consider the case when the surface of the core is covered by an elastically soft ultrathin (its thickness $\Delta R \sim (1-4)$ nm) paraelectric or high-k semiconductor shell with a screening length $\Lambda \sim 1$ nm. The coverage can be artificial (e.g., a soft organic semiconductor or vacancy-enriched $SrTiO_3$) or natural, where in the latter case it would originate from the polarization screening by surrounding media. Note that a screening length larger than 0.1 nm weakly effects the core domain structure, in this case the screening acts as an electrical open-circuit condition. Thus, we impose an electrical open-circuit condition at the cylinder ends to make the vortex-type polarization energetically favorable.

<u>Initial conditions, shape, and gradient effects variability.</u> We use a four 90-degree domain configuration in the XY-plane with superimposed small random noise as an initial distribution of polarization. These four domains determine the direction of the lateral polarization vorticity. When using a purely random noise as the initial distribution of polarization, the domain structure relaxes to structure resembling 180-degree domains in thin $c^+/c^-$ films. This process occurs because the depolarization field favors a polarization orientation along the *z*-axis of the elongated cylinder; however, the vorticity of polarization components appears near the cylinder ends (an analog of the flux-closure a-domains in thin films).

**FEM** simulations are performed in COMSOL@MultiPhysics software, using electrostatics, solid mechanics, and general math (PDE toolbox) modules. The size of the computational region is not less than 40×40×160 nm³, and is commensurate with the cubic unit cell constant (about 0.4 nm) of BaTiO$_3$ at room temperature. The minimal size of a tetrahedral element in a mesh with fine discretization is equal to the unit cell size, 0.4 nm, the maximal size is (0.8 – 1.2) nm in the core, 1.6 nm in the (1-4) nm thick elastically soft paraelectric or high-k semiconductor shell, and 4 nm in the dielectric medium (**Fig. S1**). The dependence on the mesh size is verified by increasing the minimal size to 0.8 nm. We verified that this only results in minor changes in the electric polarization, electric field, and elastic stress and strain, such that the spatial distribution of each of these quantities becomes less smooth (i.e., they contain numerical errors in the form of a small random noise). However, when using these larger cell sizes, all significant details remain visible, and more importantly, the system energy remains essentially the same with an accuracy of about 0.1%.

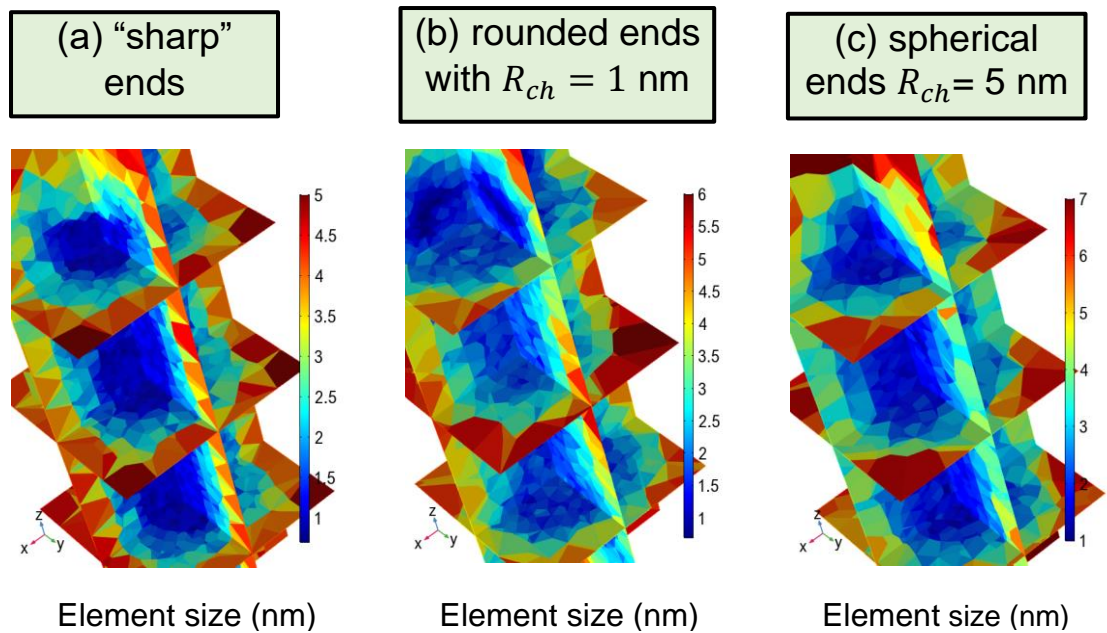

**Figure S1**. Examples of mesh for the sharp edges (a), rounded edges (b) and rounded caps (c) of the cylindrical nanoparticle. The color corresponds to the size of the mesh elements.

**Table SI.** LGD coefficients and other material parameters of BaTiO$_3$ nanocylinders

| Coefficient | Numerical value | References |
|---|---|---|
| $\varepsilon_{b,\,e}$ | $\varepsilon_b = 7$ (core background), $\varepsilon_e = 10$ (surrounding) | 14 |
| $a_i$  (in mJ/C$^2$) | $a_1 = 3.34(T-381)\cdot 10^5$, $\alpha_T = 3.34\cdot 10^5$  ($a_1 = -2.94\cdot 10^7$ at 298 K) | 19 |
| $a_{ij}$  (in m$^5$J/C$^4$) | $a_{11} = 4.69(T-393)\cdot 10^6 - 2.02\cdot 10^8$, $a_{12} = 3.230\cdot 10^8$, ($a_{11} = -6.71\cdot 10^8$ at 298 K) | 19 |
| $a_{ijk}$  (in m$^9$J/C$^6$) | $a_{111} = -5.52(T-393)\cdot 10^7 + 2.76\cdot 10^9$, $a_{112} = 4.47\cdot 10^9$, $a_{123} = 4.91\cdot 10^9$ (at 298 K $a_{111} = 82.8\cdot 10^8$, $a_{112} = 44.7\cdot 10^8$, $a_{123} = 49.1\cdot 10^8$) | 19 |
| $Q_{ij}$ (m$^4$/C$^2$) | $Q_{11}=0.11$, $Q_{12}= -0.043$, $Q_{44}=0.059$ | 19 |
| $s_{ij}$  (in $10^{-12}$ Pa$^{-1}$) | $s_{11}=8.3$, $s_{12}= -2.7$, $s_{44}=9.24$ | 20 |
| $g_{ij}$  (in $10^{-10}$m$^3$J/C$^2$) | $g_{11}=5.0$, $g_{12}= -0.2$, $g_{44}= 0.2$ | 14 |
| $F_{ij}$ (in $10^{-11}$m$^3$/C) $f_{ij}$ (in V) | $F_{11} = 2.4$, $F_{12} = 0.5$, $F_{44} = 0.06$ (these values are recalculated from the values  $f_{11} = 5.1, f_{12} = 3.3, f_{44} = 0.065$ V calculated in [21] The equality $F_{44} = F_{11} - F_{12}$ is valid in the isotropic case. | 21 |
| $v_{ijklm}$ | 0 (since its characteristic values are unknown for BaTiO$_3$ and other perovskites) | |
| $a_i^{(s)}$ | 0 (that corresponds to the so-called natural boundary conditions) | |
| $\beta_T^{(c)}$ (in $10^{-6}$K$^{-1}$) | 9.8 (thermal expansion coefficient) | 22, 23 |
| $a_{cubic}^{(c)}$ (in Å) | 4.035 Å lattice constant at 1000 °C | Recalculated from [23] |
| $R$ and $h$ (in nm) | $R = 10$ (vary from 2 to 20 nm), $h = 80$ (vary from 8 to 160 nm) | |

Effect of the cylinder ends shape on the polarization distribution

FEM, performed for the sharp and rounded ends of cylindrical core at room temperature with nominal values of the flexoelectric tensor components $F_{ij}$ (listed in **Table SI**), reveals that the distribution of the polarization component $P_3$ depends critically on the curvature radius of the cylinder ends (**Fig. S2**). The increase of the curvature leads to both a strong increase of the $P_3$ value and a simultaneous gradual transformation of the multiple edge domains into a bidomain c$^+$/c$^-$ configuration, which contains flux-closure domains near the spherical ends (top row in **Fig. S2**). For the sharp and slightly rounded ends, the distribution $P_3$ becomes much more contrasting with increasing $F_{ij}$; but the dependence of $P_3$ on $F_{ij}$ virtually disappears as the curvature increases (compare the top and bottom rows in **Fig. S2**). FEM results presented below are performed for the cylindrical core with sharp ends, because this shape is the most sensitive to the flexoelectric coupling.

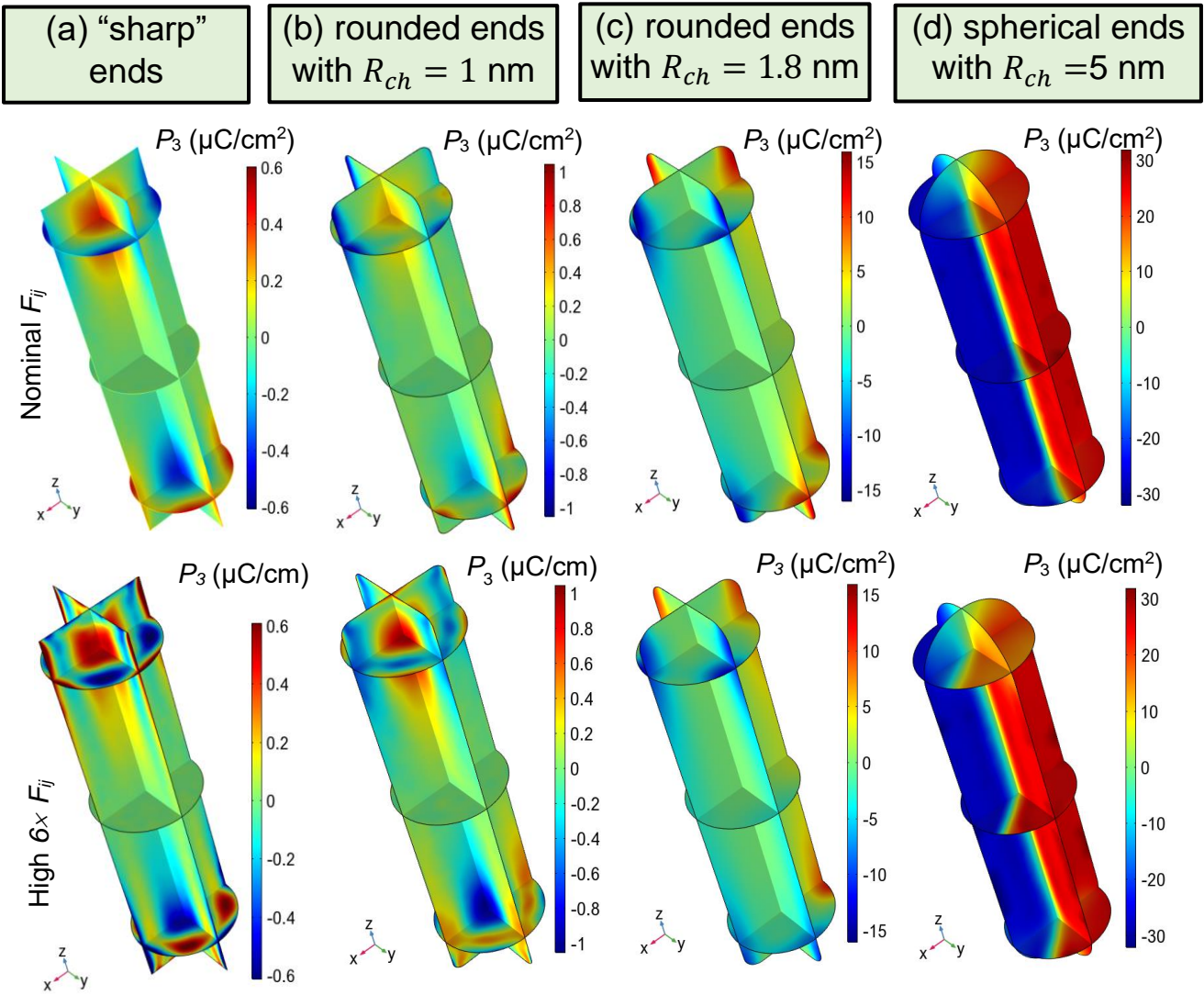

**Figure S2**. Distribution of the polarization component $P_3$ in three different cross-sections of the cylindrical nanoparticle with sharp ends **(a)**, rounded ends with curvature $R_{ch}$ =1 nm **(b)** and 1.8 nm **(c)**; and hemispherical ends **(d)**. The values of the flexoelectric coefficients are $F_{11} = 2.4 \cdot 10^{-11}$m$^3$/C, $F_{12} = 0.5 \cdot 10^{-11}$m$^3$/C, and $F_{44} = 0.06 \cdot 10^{-11}$m$^3$/C for the **top row**; $F_{11} = 144 \cdot 10^{-11}$m$^3$/C, $F_{12} = 3 \cdot 10^{-11}$m$^3$/C, and $F_{44} = 0.36 \cdot 10^{-11}$m$^3$/C for the **bottom row.** Temperature $T = 300\ K$. All other parameters are listed in **Table SI**.

Note, that the vortex stability in a ferroelectric nanoparticle depends on many physical factors including the particle geometry, electric screening and elastic boundary conditions. In particular, depending on the electric boundary conditions and aspect ratio of length to radius, the vortex-type polarization can be stable, metastable or unstable in a very prolate cylinder. Here we use not very prolate cylinders with an aspect ratio varying from 4:1 to 8:1 (see **Table SI**); and the cylinder ends are electrically open, which corresponds to a weak electric screening. The electrical open-circuit condition at the cylinder ends makes the vortex-type polarization energetically favorable independently on the presence of the flexoelectric coupling. The single-domain state can be stable in a much more prolate cylinders with an aspect ratio much more than 10 covered by the shell with a very small or zero effective screening length Λ (see e.g., Ref.[24]).

Effect of polarization gradient coefficients on the polarization distribution

We performed FEM for temperatures (200 – 400) K using zero and nominal values of the electrostriction tensor components $Q_{ij}$; and zero, positive, and negative values of $F_{ij}$. Three values of $g_{ij}$ were taken: a nominal value and a factor of ten larger and smaller than the nominal value. The nominal values of $Q_{ij}$, $F_{ij}$, and $g_{ij}$ are listed in **Table SI.** Large values of $g_{ij}$ lead to an increase of the domain wall thickness and makes the domain structure insensitive to $F_{ij}$ values; very large values of $g_{ij}$ eventually prevent the domain formation. This result expected, since, according to the Ginzburg-Landau theory, the thickness of the uncharged 180-degree domain wall is proportional to $\sqrt{g_{44}}$ for the considered geometry. The situation for small values of $g_{ij}$ is much more interesting.

Typical distributions of the polarization component $P_3$, and polarization magnitude $P$ calculated for small $g_{ij}$ and the different values of electrostriction and flexoelectric coupling coefficients are shown in **Fig. S3**. For $Q_{ij} = 0$ and $F_{ij} = 0$, the reduction of $g_{ij}$ by a factor of ten leads to the decrease of the domain wall width only (**Fig. S3a**). For nonzero nominal values of $Q_{ij}$, the reduction of $g_{ij}$ by a factor of ten leads to the appearance of quasi-periodic spot-like $P_3$-domains located near the lateral surface of the cylinder (shown by blue and red spots in the top row of **Figs. S3b-d**) instead of two diffuse axial domains shown in **Fig. S2a**. These spot-like domains are insensitive to the magnitude and sign of $F_{ij}$, if its absolute value is less than $4 \cdot 10^{-11}$m$^3$/C (**Figs. S3b-d**). The sharp transformation of the quasi-stable spot-like $P_3$-domains into a stable bidomain configuration with flux-closure domains at the cylinder ends appears at unrealistically high values of $F_{ij}$ (**Fig. S3e**). The distributions of polarization magnitude $P$ calculated for nonzero $Q_{ij}$ and $|F_{ij}| < 4 \cdot 10^{-11}$m$^3$/C reveal a twisted central line with quasi-periodically located multiple Bloch points, $\boldsymbol{P} = 0$, whose patterns are shown in the bottom row of **Figs. S3b-d**.

The influence of small $g_{ij}$ values can be explained in the following way. According to the Ginzburg-Landau theory, the thickness of the domain wall is directly associated with the gradient coefficient. Thus, a decrease in the gradient coefficients leads to a decrease in the thickness of the domain walls, which affects the size of the domains in the transverse direction. This is exactly what is shown in **Fig. S3,** calculated for small values of the gradient coefficients. The small $g_{ij}$ values leads to the decrease of the domain wall energy, and, consequently, can lead to the formation of labyrinthine domain patterns [3, 4] or other gradient-driven structural instabilities, such as domain wall meandering [25, 26]. An increase in flexoelectric coefficients leads to an increase in effective gradient coefficients, as a consequence of this, the domain size increases. More importantly, the increase of $|F_{ij}|$ to the value comparable with $\sqrt{\frac{|g_{ij}|}{c}}$, where $c$ is the geometry-dependent combination of elastic modulus, can lead to the appearance of spatially modulated domain patterns [27]. For the small $g_{ij}$ values used in **Fig. S3b-d**, the influence of $g_{ij}$ and $F_{ij}$ support the formation of spatially-modulated domain spots and Bloch points twisting (similar to meandering).

FEM performed for sharp and rounded ends of a cylindrical BaTiO$_3$ core reveals that the distribution of the polarization components depends critically on the curvature radius of the cylinder ends and is highly sensitive to the values of polarization gradient coefficients.

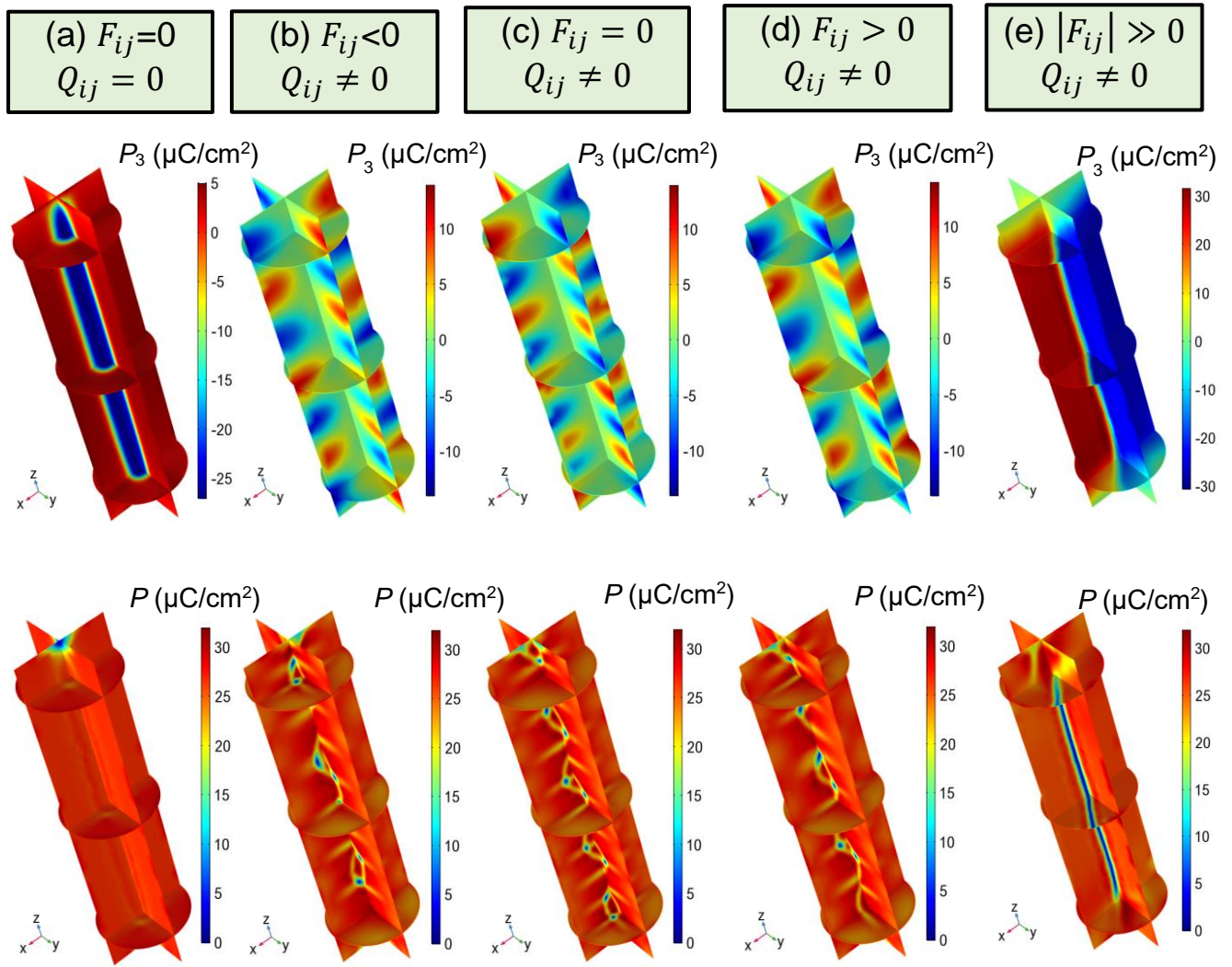

**Figure S3**. Distribution of the polarization component $P_3$ (**top row**) and polarization magnitude $P$ (**bottom row**) in three different cross-sections of the nanoparticle core. Different columns are calculated for the different values of electrostriction and flexoelectric coupling coefficients. The following values of gradient coefficients are used: $g_{11} = 1 \cdot 10^{-10}$ m$^3$/F, $g_{12} = -0.04 \cdot 10^{-10}$ m$^3$/F, and $g_{44} = 0.04 \cdot 10^{-11}$ m$^3$/F. Temperature $T = 300\,K$. All other parameters are listed in **Table SI**.

Effect of the shell dielectric properties on the domain structure

The dominant vortex-type polarization $\vec{P}_1 + \vec{P}_2$, which is directed tangentially to the lateral surface of the cylinder (**Fig. S4**, top row), appears much bigger than the axial component $P_3$ perpendicular to the cylinder ends. Actually, a comparison of the scales for $P_3$ and $P$ in **Fig. S4** shows that $P \gg |P_3|$ and $P \approx |\vec{P}_1 + \vec{P}_2|$, and so $|\vec{P}_1 + \vec{P}_2| \gg |P_3|$. Consequently, the resulting depolarization field $E_3$ created by the component $P_3$ is very small inside the particle. The field $E_3$ is determined by the dielectric properties of the shell and, in particular, $E_3$ slightly decreases with the increase of the shell dielectric permittivity. A shell with a very high relative dielectric permittivity (e.g., vacancy-enriched paraelectric SrTiO$_3$) can provide an effective screening of the axial polarization component in a prolate cylindrical multiaxial ferroelectric core, while the dielectric properties of the shell virtually do not effect on vortex-type polarization $\vec{P}_1 + \vec{P}_2$. Being energetically preferable, the appearance of a shell-insensitive vortex-type polarization in a multiaxial ferroelectric core, such as BaTiO$_3$, is very probable [28]. This qualitative statement is confirmed quantitatively by the very small $div\vec{P}$ and depolarization electric field $E_3$ in the core (**Fig. S9a** and **S9c**).

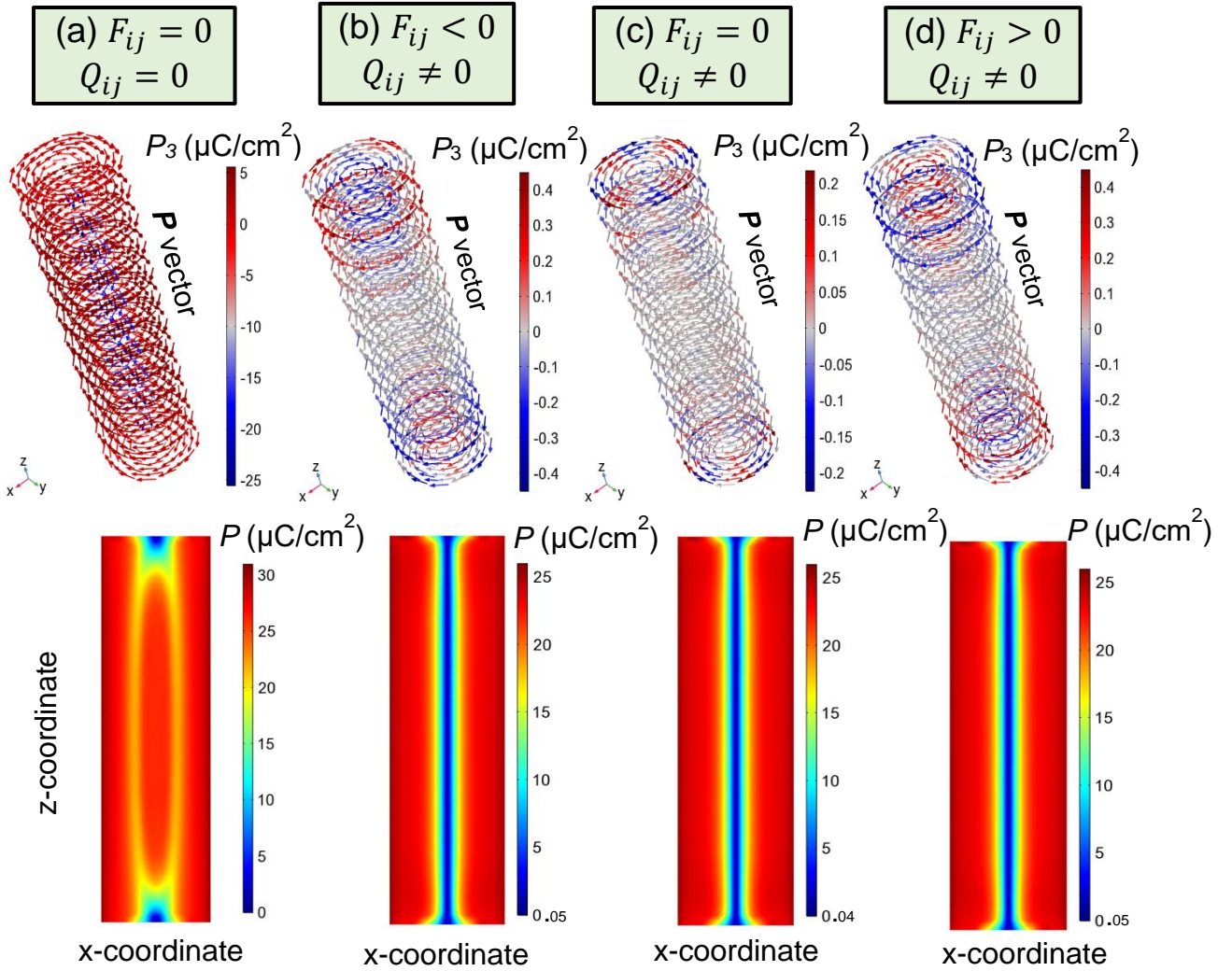

**Figure S4**. Distribution of the polarization vector **P** inside a cylindrical nanoparticle (**top row**) and polarization magnitude $P$ (**bottom row**) in the XZ cross-section of the nanoparticle. The images are calculated without electrostriction ($Q_{ij} = 0$) and flexoelectric ($F_{ij} = 0$) couplings (**a**). In the case where electrostriction coupling in non-zero ($Q_{ij} \neq 0$), the values of the flexoelectric coefficients $F_{ij}$ are varied: negative (**b**), zero (**c**), and positive (**d**). The arrow color corresponds to the value of $P_3$ (red-blue scale). The values of $F_{ij}$ and all other parameters are given in **Table SI**, $T = 300$ K.

A strong influence of the shell dielectric properties on the core domain structure is possible for a uniaxial ferroelectric core [29], since a polarization rotation is impossible in this case. In particular, a weak screening can lead to the formation of labyrinthine domains in nanoparticles of uniaxial ferroelectrics $CuInP_2S_6$ [3] and $Sn_2P_2S_6$ [4].

The elastic properties of the shell are important for spherical core-shell nanoparticles, where the vortices orientation is determined by an elastic anisotropy of the shell [30]. FEM calculations show that in the case of the cylindrical core-shell $BaTiO_3$ nanoparticles, the vortex axis is oriented along the cylinder axis and is nearly independent of the shell dielectric and elastic properties; therefore, role of the shell for a prolate multiaxial cylindrical core is much weaker than it is for a uniaxial spherical core.

**APPENDIX B. Influence of the Flexoelectric Coupling Strength and Anisotropy on the Domain Structure of a Cylindrical Nanoparticle**

The distribution of the polarization component $P_3$ in three different cross-sections of the nanoparticle core, polarization magnitude $P$ in the XZ cross-section of the nanoparticle, and isosurfaces of polarization components $P_{1,2,3}$ are shown in **Fig. S5**. The images are calculated without electrostriction ($Q_{ij} = 0$) and flexoelectric ($F_{ij} = 0$) coupling (**Fig. S5a**); with electrostriction coupling ($Q_{ij} \neq 0$) and negative (**Fig. S5b**), zero (**Fig. S5c**), positive (**Fig. S5d**), or high positive (**Fig. S5e**) values of the flexoelectric coefficients $F_{ij}$.

The extremal (maximal or minimal) value $P_e$ in the center of the diffuse axial $P_3$-domain, shown by the green curve in **Fig. 3f** (main text), frequently differs from $P_{max}$ and $P_{min}$ values due to the presence of the small sixteen $P_3$-domains localized near the top and bottom junction of cylindrical sidewall with the ends (bottom raw in **Fig. S5**). The polarization direction alternates in these 180-degree domains, with four "up" and four "down" domains at each junction. These alternating $P_3$-domains are not the part of flexon, because they are induced by the electrostriction coupling. Their structure depends weakly on the flexoelectric coupling strength $f$, if $f$ is small. However, the shape of each domain significantly changes and the size of each domain moderately increases as $|F_{ij}|$ increases (compare **Fig. S5d** and **Fig. S5e**). The change of polarization direction occurs in each of the 16 alternating $P_3$-domains at $F_{ij} \to -F_{ij}$ (compare **Fig. S5b** and **Fig. S5d**).

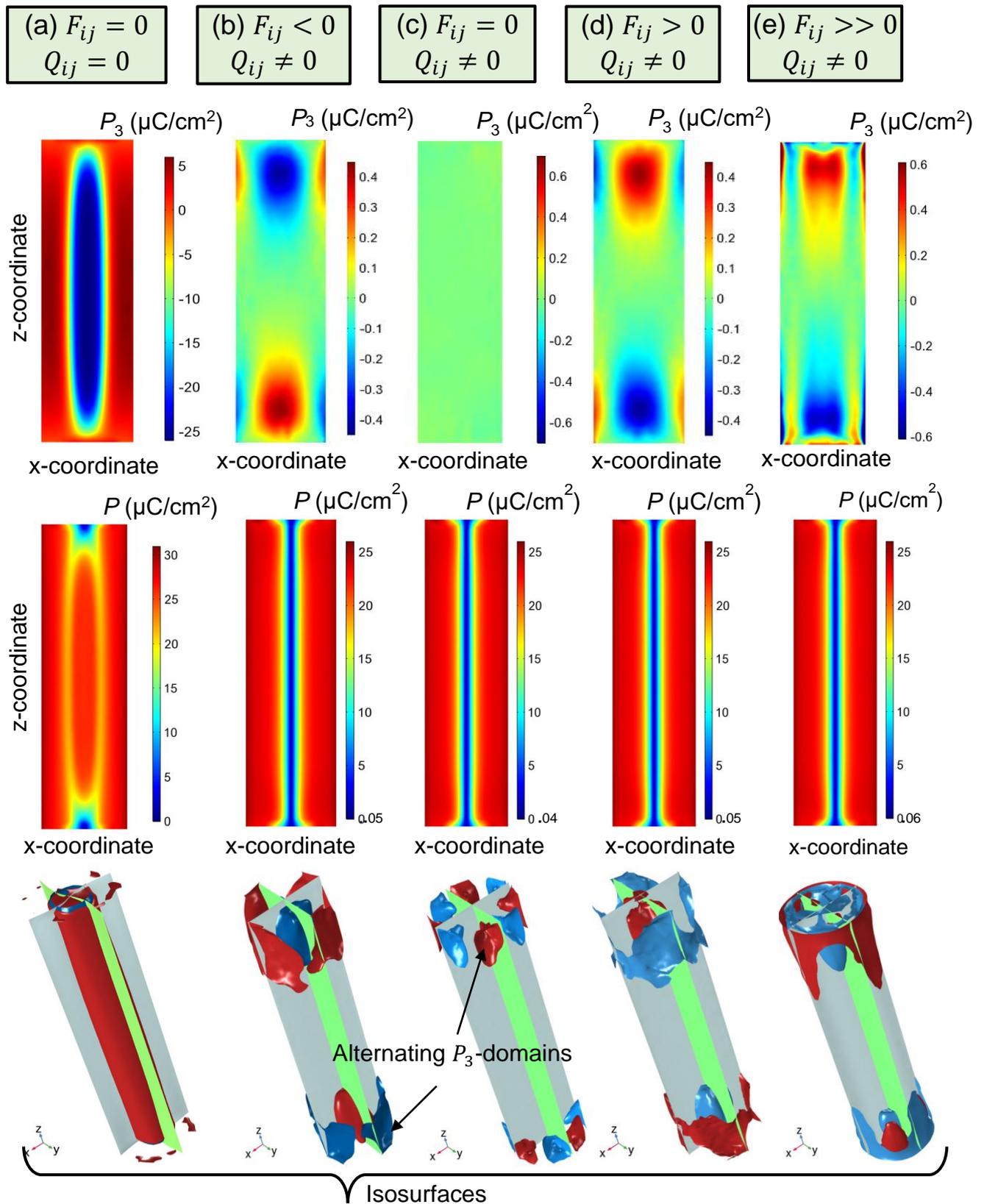

**Figure S5**. Distribution of the polarization component P$_3$ (**top row**) and polarization magnitude (**middle row**) in the XZ cross-section of the nanoparticle. **Bottom row**: isosurfaces of zero polarization components $P_1$ (gray), $P_2$ (green), and nonzero $P_3 = \pm 0.1, \pm 0.1, \pm 0.2$, and $\pm 0.4$ µC/cm² (red and blue). Images are calculated without electrostriction ($Q_{ij} = 0$) and flexoelectric ($F_{ij} = 0$) coupling (**a**); with electrostriction coupling ($Q_{ij} \neq 0$) and negative (**b**), zero (**c**), positive (**d**), or tripled positive (**e**) values of flexoelectric coefficients $F_{ij}$. Reference values of $F_{ij}$ and all other parameters are given in **Table SI**. Temperature $T = 300\ K$.

The distribution of the polarization component $P_3$ in three different cross-sections of the nanoparticle core, $P_3$ distribution XZ cross-section, and isosurfaces of polarization components $P_{1,2,3}$ are shown in **Fig. S6**. Images are calculated for the fixed values $F_{11} = 2.4 \cdot 10^{-11} m^3/C$ and $F_{12} = 0.5 \cdot 10^{-11} m^3/C$ are, and the value of $F_{44}$ varies from -6 to 6 (in $10^{-11} m^3/C$) as indicated in the legends to **Fig. S6a-f**. The flexon containing two pronounced axial domains located near the cylinder ends exists at large negative (**Fig. S6a**) and large positive (**Fig. S6f**) $F_{44}$ values. The $P_3$-domains become smaller and more diffuse with a decrease of $|F_{44}|$; but they are still visible and practically do not change their shape, size, or polarization distribution for small $|F_{44}|$ values over the range $|F_{44}| \leq 0.06$ (**Fig. S6b-d**). The flexon becomes faint and almost disappears when $F_{44}$ approaches the value $F_{44} = F_{11} - F_{12} = 1.9 \cdot 10^{-11}\ m^3/C$, corresponding to the isotropic symmetry of $F_{ij}$ (**Fig. S6e**).

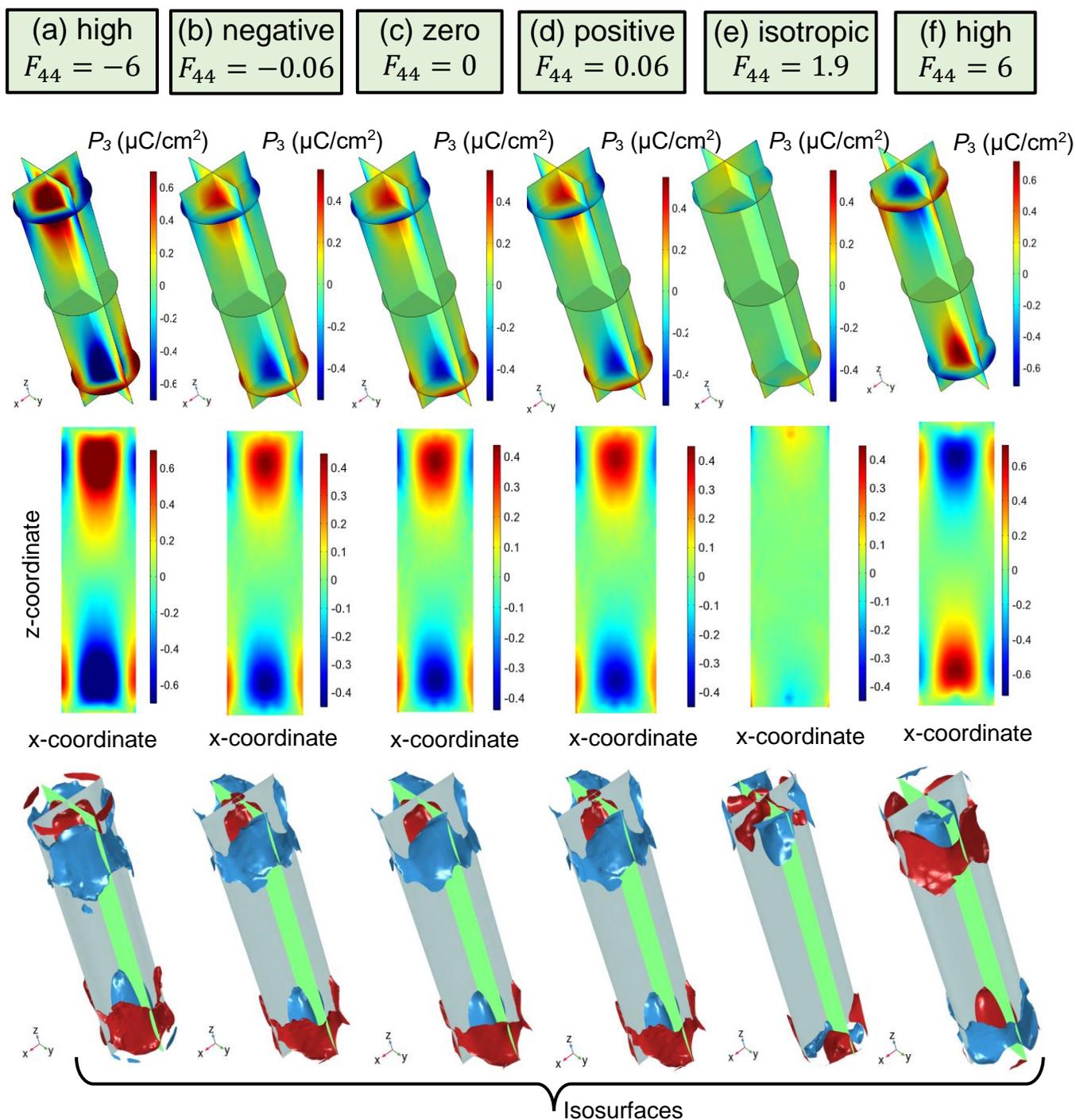

**Figure S6**. Distribution of polarization component $P_3$ in three different cross-sections of the nanoparticle core (**top row**), and $P_3$ in the XZ cross-section of cylindrical core (**middle row**). **Bottom row:** isosurfaces of polarization components $P_1$ (gray), $P_2$ (green), and $P_3$ (red and blue). Zero values correspond $P_1$ and $P_2$ isosurfaces, while isosurfaces of $P_3$ correspond to the values $P_3 = \pm 0.1, \pm 0.2,$ and $\pm 0.4$ μC/cm² for the images **(a)-(d)**, **(e),** and **(f)** respectively. The images are calculated for the fixed values $F_{11} = 2.4 \cdot 10^{-11}$m³/C and $F_{12} = 0.5 \cdot 10^{-11}$m³/C, while the value of $F_{44}$ varies from -6 to 6 (in $10^{-11}$m³/C) as indicated in the legends. All other parameters are listed in **Table SI**. Temperature $T = 300\ K$.

**APPENDIX C. Temperature Behavior of the Flexon-Type Polarization Distribution**

To define the temperature interval, where flexons exist being stable or meta-stable, we performed FEM in the temperature range from 50 K to 400 K using different initial distributions of polarization in a cylindrical core. Typical FEM results are shown in **Fig. S7**, where the columns (a)-(e) correspond to the temperature increase from 240 K to 370 K.

A bidomain configuration of $P_3$ is stable at temperatures lower than 250 K (**Figs. S7a**). The bidomain structure has a relatively thin uncharged 180-degree domain wall inside the cylinder, which transforms into a flux-closure domain near the electrically-open cylinder ends (top and middle images in **Figs. S7a**). The origin of the flux-closure domain is the core tendency to minimize its electrostatic energy, because a flux-closure domain wall creates a much weaker depolarization field (in fact negligibly small) compared to the field that the charged 180-degree domain wall would create. The structure of the P-vector is vortex-like and changes moderately when approaching the surface at $T \leq$ 240 K (as seen from the direction of arrows in the bottom image of **Figs. S7a**).

We revealed that an initial four-domain polarization distribution relaxes to a flexon-like domain structure in the temperature range $260\text{ K} < T < 360\text{ K}$ (middle row in **Figs. S7b-d**). The structure of the P-vector is vortex-like, which weakly changes when approaching the surface in the same temperature range (as seen from the direction of arrows at the bottom image of **Figs. S7b-d**). A relatively small domain wall broadening exists near the cylinder ends, which can be seen from the polarization magnitude distribution at $260\text{ K} < T < 360\text{ K}$ (top images of **Figs. S7b-d**).

The flexon gradually disappears at $T > 370$ K (middle image in **Figs. S7e**). The ferroelectric polarization inside the core significantly decreases at $T > 370$ K and completely disappears at $T \sim 400$ K (top image in **Figs. S7e**). The lateral components $P_1$ and $P_2$ form a vortex-like structure and their distribution is nearly independent of the coordinate $z$ along the cylinder axis. Hence, the vortex-like structure of the P-vector is insensitive to the surface presence. The structure becomes faint with a temperature increase above 370 K (note that the arrows length decreases in the bottom image of **Figs. S7e**). The flexon-type polarization distribution exists over a relatively wide temperature range $260\text{ K} < T < 360\text{ K}$. The axial counter domains inherent to flexons are the most pronounced feature over the narrower range $290\text{ K} < T < 340\text{ K}$. The relatively wide temperature range (about 100 K) corresponding to the stability or meta-stability of the flexon-type polarization distribution gives us the hope that the domain morphology can be observed experimentally.

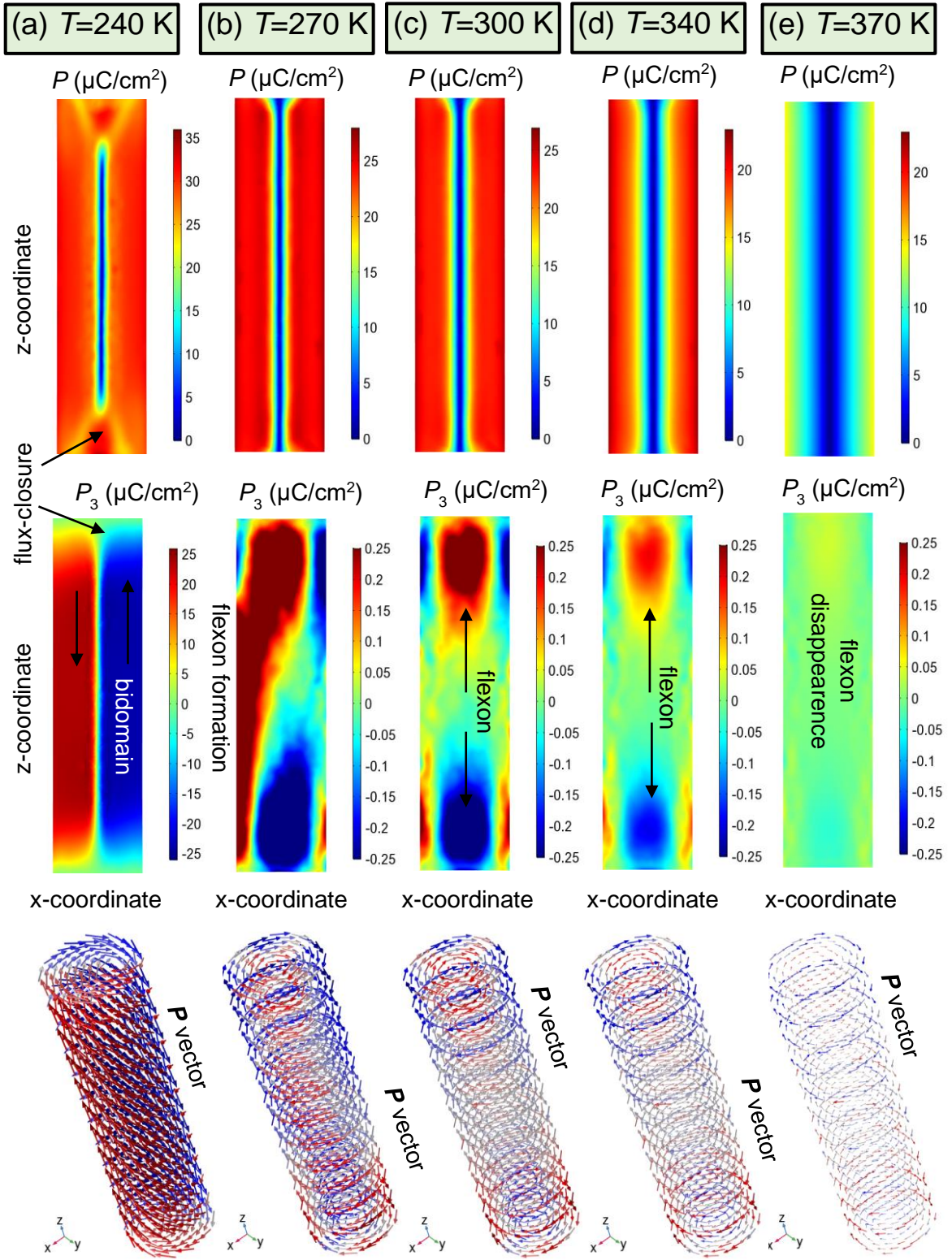

**Figure S7**. Distribution of polarization magnitude $P$ (**top row**) and its component $P_3$ (**middle row**) in XZ cross-sections of the nanoparticle core. **Bottom row:** distribution of polarization vector **P** shown by arrows colored according to the value of $P_3$. Different columns are calculated for the temperatures $T = 240, 270, 300, 340,$ and 370 K **(a, b, c, d, e)**. All other parameters are listed in **Table SI**.

# APPENDIX D. Influence of an External Electric Field on the Flexon Dynamics

The orientation of $P_3$ can be readily changed by the application of an external electric field. Typical quasistatic hysteresis loops of $P_3(U)$ are shown in **Figs. S8a-d**, where the black loops correspond to the average polarization $\overline{P_3}(U)$; blue and red loops correspond to the minimal ($P_{min}(U)$) and maximal ($P_{max}(U)$) values of $P_3(U)$ at a given voltage $U$. The loops of $\overline{P_3}(U)$ are very slim, voltage-symmetric (since their shape is symmetrical with respect to the transformation $U \to -U$), and belong to the antiferroelectric type (since they are double loops with $\overline{P_3}(U) = 0$). The loops of $P_{min}(U)$ and $P_{max}(U)$ are strongly voltage-asymmetric and contain two single loops of different shape and size, which we refer to as "major" and "minor" loops. The loop $P_{min}(U)$ is strongly shifted downward, and the loop $P_{max}(U)$ is strongly shifted upward.

The loops in **Figs. S8a-S8c** are calculated for nonzero electrostriction coupling coefficients $Q_{ij}$ and negative, zero, or positive flexoelectric coupling coefficients $F_{ij}$. The main difference between the loops shown in **Figs. S8a, S8b,** and **S8c**, appear in a horizontal shift of the asymmetric loops $P_{min}(U)$ and $P_{max}(U)$, originating from the effective "flexoelectric" field. The field is proportional to the combination of $F_{ij}$, and so its direction is defined by the sign of $F_{ij}$, and it is absent for $F_{ij} = 0$. In the linear approximation, the flexoelectric field is proportional to $(F_{11} - F_{44} - F_{12})\frac{\partial u_{33}}{\partial z}$, where $u_{33}$ is the component of elastic strain tensor (details are in **Appendix E**). The proportionality partially explains the right shift ($U_{int} > 0$) of the intersection point between major and minor loops calculated for $F_{ij} \leq 0$ (shown in **Figs. S8a-b**) and the left shift of the intersection point calculated for $F_{ij} > 0$ (shown in **Fig. S8c**).

The loops in **Fig. S8d** are calculated without electrostriction ($Q_{ij} = 0$) and flexoelectric coupling ($F_{ij} = 0$). Here the loop of $\overline{P_3}(U)$ is almost indistinguishable from the loops $\overline{P_3}(U)$ shown in **Figs. S8a-8c** for $Q_{ij} \neq 0$. The subtle difference is an ultra-small coercivity at $U = 0$, which is absent for the loops in **Figs. S8a-c**. The minor loops of $P_{min}(U)$ and $P_{max}(U)$ are very small in comparison with the major loops, and they are also significantly smaller than the minor loops shown in **Figs. S8a-S8c**. The intersection point of major and minor loops corresponds to $U_{int} = 0$.

The bottom row in **Fig. S8** shows $P_3$ distributions for different voltages, from which it follows that different $P_3$ distributions averaged over the core volume correspond to the same average value, $\overline{P_3}(U)$. Thus, the analysis of the hysteresis loops leads to the conclusion that flexons cannot be distinguished from macroscopic (e.g., capacitance) measurements of the average polarization in a homogeneous electric field, they can only be registered by local probing methods using the strong gradient of electric field with a nanoscale resolution, such as piezoresponse force microscopy (**PFM**).

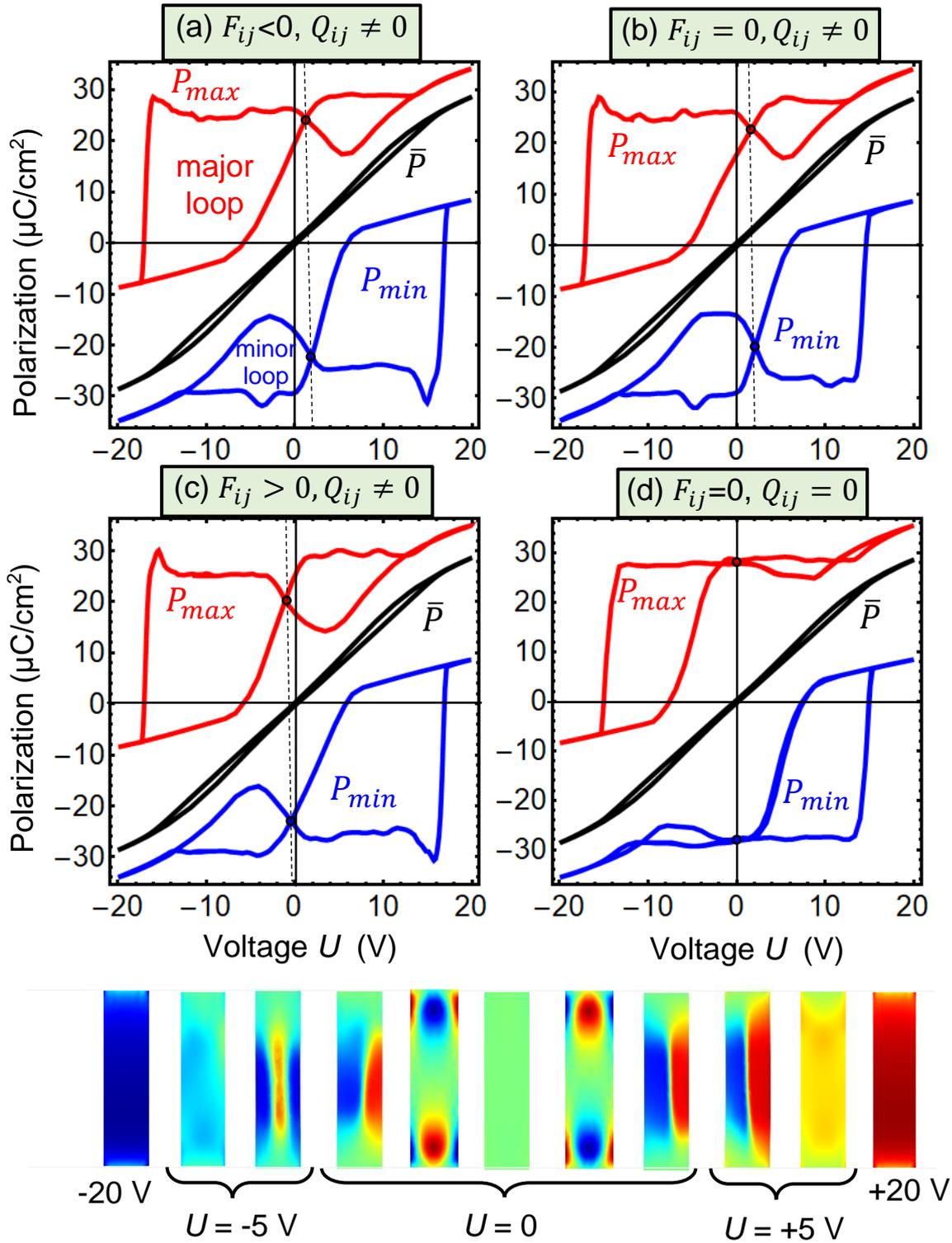

**Figure S8.** Dependence of the average polarization component $\overline{P_3}$ (black loops) and its minimal $P_{min}$ (blue loops) and maximal $P_{max}$ (red loops) values on the applied voltage $U$. Black dots connected with a vertical dashed line correspond to $U_{int}$. Plots **(a-d)** are calculated for different values of electrostriction and flexoelectric coupling coefficients, as written in the legends. The plots **(a-c)** are calculated for nonzero electrostriction coupling coefficients ($Q_{ij} \neq 0$) and negative **(a)**, zero **(b)**, or positive **(c)** flexoelectric coupling coefficients $F_{ij}$. Plot **(d)** is calculated without electrostriction ($Q_{ij} = 0$) and flexoelectric ($F_{ij} = 0$) coupling. Temperature $T = $

300 $K$, the distance between the electrodes is 60 nm. All other parameters are listed in **Table SI**. The bottom row shows $P_3$ distributions for different voltages $U$. Different distributions corresponding to the same $U$ illustrate a rather small hysteresis effect.

## APPENDIX E. Approximate Analytical Solution

Since $\frac{\partial \sigma_{i1}}{\partial x_i} = \frac{\partial \sigma_{i2}}{\partial x_i} = \frac{\partial \sigma_{i3}}{\partial x_i} = 0$ in accordance with mechanical equilibrium equations, the right-hand sides in Eqs. (A.3) can be rewritten as:

$$\Gamma \frac{\partial P_1}{\partial t} + 2P_1(a_1 - Q_{12}(\sigma_{22} + \sigma_{33}) - Q_{11}\sigma_{11}) - Q_{44}(\sigma_{12}P_2 + \sigma_{13}P_3) + 4a_{11}P_1^3 + 2a_{12}P_1(P_2^2 + P_3^2) + 6a_{111}P_1^5 + 2a_{112}P_1(P_2^4 + 2P_1^2 P_2^2 + P_3^4 + 2P_1^2 P_3^2) + 2a_{123}P_1 P_2^2 P_3^2 - g_{11}\frac{\partial^2 P_1}{\partial x_1^2} - g_{44}\left(\frac{\partial^2 P_1}{\partial x_2^2} + \frac{\partial^2 P_1}{\partial x_3^2}\right) = -(F_{11} - F_{44} - F_{12})\frac{\partial \sigma_{11}}{\partial x_1} - F_{12}\frac{\partial}{\partial x_1} Tr\hat{\sigma} + E_1 \quad (E.1a)$$

$$\Gamma \frac{\partial P_2}{\partial t} + 2P_2(a_1 - Q_{12}(\sigma_{11} + \sigma_{33}) - Q_{11}\sigma_{22}) - Q_{44}(\sigma_{12}P_1 + \sigma_{23}P_3) + 4a_{11}P_2^3 + 2a_{12}P_2(P_1^2 + P_3^2) + 6a_{111}P_2^5 + 2a_{112}P_2(P_1^4 + 2P_2^2 P_1^2 + P_3^4 + 2P_2^2 P_3^2) + 2a_{123}P_2 P_1^2 P_3^2 - g_{11}\frac{\partial^2 P_2}{\partial x_2^2} - g_{44}\left(\frac{\partial^2 P_2}{\partial x_1^2} + \frac{\partial^2 P_2}{\partial x_3^2}\right) = -(F_{11} - F_{44} - F_{12})\frac{\partial \sigma_{22}}{\partial x_2} - F_{12}\frac{\partial}{\partial x_2} Tr\hat{\sigma} + E_2 \quad (E.1b)$$

$$\Gamma \frac{\partial P_3}{\partial t} + 2P_3(a_1 - Q_{12}(\sigma_{11} + \sigma_{22}) - Q_{11}\sigma_{33}) - Q_{44}(\sigma_{13}P_1 + \sigma_{23}P_2) + 4a_{11}P_3^3 + 2a_{12}P_3(P_1^2 + P_2^2) + 6a_{111}P_3^5 + 2a_{112}P_3(P_1^4 + 2P_3^2 P_1^2 + P_2^4 + 2P_2^2 P_3^2) + 2a_{123}P_3 P_1^2 P_2^2 - g_{11}\frac{\partial^2 P_3}{\partial x_3^2} - g_{44}\left(\frac{\partial^2 P_3}{\partial x_1^2} + \frac{\partial^2 P_3}{\partial x_2^2}\right) = -(F_{11} - F_{44} - F_{12})\frac{\partial \sigma_{33}}{\partial x_3} - F_{12}\frac{\partial}{\partial x_3} Tr\hat{\sigma} + E_3 \quad (E.1c)$$

The derivation is straightforward: $-F_{11}\frac{\partial \sigma_{33}}{\partial x_3} - F_{12}\left(\frac{\partial \sigma_{11}}{\partial x_3} + \frac{\partial \sigma_{22}}{\partial x_3}\right) - F_{44}\left(\frac{\partial \sigma_{13}}{\partial x_1} + \frac{\partial \sigma_{23}}{\partial x_2}\right) = -(F_{11} - F_{44})\frac{\partial \sigma_{33}}{\partial x_3} - F_{12}\left(\frac{\partial \sigma_{11}}{\partial x_3} + \frac{\partial \sigma_{22}}{\partial x_3}\right) = -(F_{11} - F_{44} - F_{12})\frac{\partial \sigma_{33}}{\partial x_3} - F_{12}\frac{\partial}{\partial x_3} Tr\hat{\sigma}$, where $Tr\hat{\sigma} = \sigma_{11} + \sigma_{22} + \sigma_{33}$ and $\frac{\partial \sigma_{13}}{\partial x_1} + \frac{\partial \sigma_{23}}{\partial x_2} \equiv -\frac{\partial \sigma_{33}}{\partial x_3}$.

Elastic stresses existing in the system can be found from Eq.(A.6) as:

$$\sigma_{11} = c_{11}u_{11} + c_{12}(u_{22} + u_{33}) - q_{11}P_1^2 - q_{12}(P_2^2 + P_3^2) - f_{11}\frac{\partial P_1}{\partial x_1} - f_{12}\left(\frac{\partial P_2}{\partial x_2} + \frac{\partial P_3}{\partial x_3}\right), \quad (E.2a)$$

$$\sigma_{22} = c_{11}u_{22} + c_{12}(u_{11} + u_{33}) - q_{11}P_2^2 - q_{12}(P_1^2 + P_3^2) - f_{11}\frac{\partial P_2}{\partial x_2} - f_{12}\left(\frac{\partial P_1}{\partial x_1} + \frac{\partial P_3}{\partial x_3}\right), \quad (E.2b)$$

$$\sigma_{33} = c_{11}u_{33} + c_{12}(u_{11} + u_{22}) - q_{11}P_3^2 - q_{12}(P_2^2 + P_1^2) - f_{11}\frac{\partial P_3}{\partial x_3} - f_{12}\left(\frac{\partial P_1}{\partial x_1} + \frac{\partial P_2}{\partial x_2}\right) \quad (E.2c)$$

$$\sigma_{12} = c_{44}u_{12} - q_{44}P_1P_2 - f_{44}\left(\frac{\partial P_1}{\partial x_2} + \frac{\partial P_2}{\partial x_1}\right), \quad \text{(E.2d)}$$

$$\sigma_{13} = c_{44}u_{13} - q_{44}P_1P_3 - f_{44}\left(\frac{\partial P_1}{\partial x_3} + \frac{\partial P_3}{\partial x_1}\right), \quad \text{(E.2e)}$$

$$\sigma_{23} = c_{44}u_{23} - q_{44}P_3P_2 - f_{44}\left(\frac{\partial P_2}{\partial x_3} + \frac{\partial P_3}{\partial x_2}\right). \quad \text{(E.2f)}$$

Here $q_{44} = \frac{Q_{44}}{s_{44}}$, $c_{44} = \frac{1}{s_{44}}$ and $f_{44} = \frac{F_{44}}{s_{44}}$.

In the virtual absence of a depolarization field, which is true with high accuracy for a vortex type polarization, the divergence $div\vec{P} = \frac{\partial P_i}{\partial x_i}$ should be very small. Indeed, the approximation $div\vec{P} \approx 0$ is valid inside the cylindrical core (**Fig. S9b**).

We also note the polarization magnitude $P = \sqrt{P_1^2 + P_2^2 + P_3^2}$ is very close to the constant value everywhere, except for the vortex core (**Fig. S5, middle row**), and its derivative $\frac{\partial P}{\partial x_3}$ is negligibly small everywhere, except in the immediate vicinity of the vortex core when in contact with the cylinder ends. For this case we obtain that:

$$\frac{\partial Tr\hat{\sigma}}{\partial x_3} = \frac{\partial}{\partial x_3}\left[(c_{11} + 2c_{12})Tr\hat{u} - (q_{11} + 2q_{12})P^2 - (f_{11} + 2f_{12})div\vec{P}\right] \approx (c_{11} + 2c_{12})\frac{\partial Tr\hat{u}}{\partial x_3} \quad \text{(E.3)}$$

and

$$\frac{\partial \sigma_{33}}{\partial x_3} = \frac{\partial}{\partial x_3}\left[c_{11}u_{33} + c_{12}(u_{11} + u_{22}) + (q_{11} - q_{12})P_3^2 - q_{12}P^2 - (f_{11} - f_{12})\frac{\partial P_3}{\partial x_3} - f_{12}div\vec{P}\right] \approx$$

$$\frac{\partial}{\partial x_3}\left[(c_{11} - c_{12})u_{33} + c_{12}Tr\hat{u} + (q_{11} - q_{12})P_3^2\right] - (f_{11} - f_{12})\frac{\partial^2 P_3}{\partial x_3^2} \quad \text{(E.4)}$$

Assuming that $\frac{\partial Tr\hat{u}}{\partial x_3} \approx 0$ and using the smallness of $P_3$, we can obtain the approximate linearized equation for $P_3$:

$$2a_1^*P_3 - g_{11}^*\frac{\partial^2 P_3}{\partial x_3^2} - g_{44}\left(\frac{\partial^2 P_3}{\partial x_1^2} + \frac{\partial^2 P_3}{\partial x_2^2}\right) \approx Q_{44}(\sigma_{13}P_1 + \sigma_{23}P_2) - (F_{11} - F_{44} - F_{12})(c_{11} - c_{12})\frac{\partial u_{33}}{\partial x_3} + E_3, \quad \text{(E.5)}$$

where $a_1^* = a_1 - Q_{12}Tr\hat{\sigma} - (Q_{11} - Q_{12})\sigma_{33} \approx a_1 - Q_{12}(c_{11} + 2c_{12})Tr\hat{u} - (q_{11} + 2q_{12})p^2 - (Q_{11} - Q_{12})[(c_{11} - c_{12})u_{33} + c_{12}Tr\hat{u}] \approx a_1 - (q_{11} + 2q_{12})p^2 - (Q_{11} - Q_{12})(c_{11} - c_{12})u_{33}$ and $g_{11}^* = g_{11} + (F_{11} - F_{44} - F_{12})(f_{11} - f_{12})$. The simplest 2D polarization vortex can be modeled by the functions

$$P_1 = p(r)sin\varphi, \; P_2 = -p(r)cos\varphi \text{ and } P_3 = 0. \quad \text{(E.6a)}$$

These polarization components in cylindrical coordinates are

$$P_r = 0, \quad P_\varphi = -p(r) \text{ and } P_3 = 0. \tag{E.6b}$$

It is easy to check that in the case $div\vec{P} = 0$ for arbitrary $p(\rho)$. For a 2D-vortex with an empty core we can assume that $p(\rho) \approx tanh\left(\frac{\rho}{L_C^x}\right)$, where $L_C^x$ is a transverse correlation length.

Using Eqs.(E.6) as a zero approximation we obtain from Eq.(E.2) that $\sigma_{13} = c_{44}u_{13}$, $\sigma_{23} = c_{44}u_{23}$ and $\sigma_{33} = (c_{11} - c_{12})u_{33} + c_{12}\text{Tr}\hat{u} - q_{12}p^2$

$$P_3 \approx \frac{Q_{44}c_{44}p(u_{13}\sin\varphi - u_{23}\cos\varphi) - (F_{11} - F_{44} - F_{12})(c_{11} - c_{12})\frac{\partial u_{33}}{\partial x_3}}{2[a_1 - (q_{11} + 2q_{12})p^2 - (Q_{11} - Q_{12})(c_{11} - c_{12})u_{33} + (g_{11} + (F_{11} - F_{44} - F_{12})(f_{11} - f_{12}))L_C^z + g_{44}L_C^x]}, \tag{E.7a}$$

Next, we consider several model cases. The first case is an absolutely rigid shell covering the ferroelectric core, when the maximal stresses evolved in the system can be roughly estimated from Eqs.(A.8) at $u_{ij} = 0$. The second case corresponds to a very soft shell covering the rigid ferroelectric core, when the stresses are minimal and the strains (at least near the cylinder ends) can be roughly estimated from Eqs.(E.2) at $\sigma_{ij} = 0$. Furthermore, the compatibility conditions should be valid.

Using the relations $q_{11} - q_{12} = (Q_{11} - Q_{12})(c_{11} - c_{12}) = \frac{Q_{11} - Q_{12}}{s_{11} - s_{12}}$, $q_{11} + 2q_{12} = (Q_{11} + 2Q_{12})(c_{11} + 2c_{12}) = \frac{Q_{11} + 2Q_{12}}{s_{11} + 2s_{12}}$, $c_{11} - c_{12} = \frac{1}{s_{11} - s_{12}}$, $c_{44} = \frac{1}{s_{44}}$ and $f_{11} - f_{12} = (F_{11} - F_{12})(c_{11} - c_{12}) = \frac{F_{11} - F_{12}}{s_{11} - s_{12}}$, we obtain that

$$P_3 \approx \frac{\frac{Q_{44}}{s_{44}}p(u_{13}\sin\varphi - u_{23}\cos\varphi) - \frac{F_{11} - F_{44} - F_{12}}{s_{11} - s_{12}}\frac{\partial}{\partial z}u_{33}}{2\left[a_1 - \frac{Q_{11} + 2Q_{12}}{s_{11} + 2s_{12}}p^2 - \frac{Q_{11} - Q_{12}}{s_{11} - s_{12}}u_{33} + \left[g_{11} + (F_{11} - F_{44} - F_{12})\frac{F_{11} - F_{12}}{s_{11} - s_{12}}\right]L_C^z + g_{44}L_C^x\right]}, \tag{E.7b}$$

where $\{\rho, \varphi, z\}$ are cylindrical coordinates, the function $p(\rho) \approx tanh\left(\frac{\rho}{L_C^x}\right)$, and $L_C^x$ and are lateral and axial correlation lengths. The functions $u_{ij}(\rho, \varphi, z)$ are elastic strains, $s_{ij}$ are elastic compliances, $Q_{ij}$ are electrostriction tensor components, $g_{ij}$ are polarization gradient coefficients written in Voight notations. From Eq.(E.7b) the axial part of the flexon polarization is proportional to $-\frac{F_{11} - F_{44} - F_{12}}{s_{11} - s_{12}}\frac{\partial}{\partial z}u_{33}(\rho, \varphi, z)$, and this proportionality along with **Fig. S9o** qualitatively describes the curves' behavior in **Fig. 3f** and **4f** in the main text**.**

Using Eqs.(E.1), we can "recover" an analog of Eq.(A.2) for a "full" Lifshitz invariant:

$$G_{flexo} = \frac{F_{11} - F_{44} - F_{12}}{2}\left(\sigma_{ii}\frac{\partial P_i}{\partial x_i} - P_i\frac{\partial \sigma_{ii}}{\partial x_i}\right) + \frac{F_{12}}{2}\left[Tr(\hat{\sigma})\frac{\partial P_i}{\partial x_i} - P_i\frac{\partial Tr(\hat{\sigma})}{\partial x_i}\right], \tag{E.8a}$$

Assuming that $\frac{\partial P_i}{\partial x_i} \approx 0$ (as it should be for the mostly uncharged domain structures) and making an integration over parts in the second term in Eq.(E.8a), we obtain:

$$G_{flexo} \approx \frac{F_{11}-F_{44}-F_{12}}{2}\left(\sigma_{ii}\frac{\partial P_i}{\partial x_i} - P_i \frac{\partial \sigma_{ii}}{\partial x_i}\right). \qquad (E.8b)$$

Elementary, but cumbersome calculations lead to an "odd" flexo-field in the boundary conditions $\int_V G_{flexo} \approx \int_V G_0 + \frac{F_{11}-F_{44}-F_{12}}{2}\frac{Q_{11}-Q_{12}}{3}\int_{Si} P_i^3$.

To verify the analogy with a Dyzaloshinskii-Moryia interaction (DMI), we attempt to convert the Lifshitz invariants into chiral interactions of a DMI type. Using $div\vec{P} = 0$ for the uncharged domain structures and making straightforward analytic manipulations for the cubic m3m point symmetry group of the BaTiO$_3$ parent phase, the explicit form of the Lifshitz invariant (1f) is $G_{flexo} \approx \frac{F_{11}-F_{44}-F_{12}}{2}\left(\sigma_{ii}\frac{\partial P_i}{\partial x_i} - P_i \frac{\partial \sigma_{ii}}{\partial x_i}\right)$, where a summation over "$i$" is performed [Eqs.(E.8)]. The elastic stress $\sigma_{ij}$ contains a contribution proportional to $Q_{ijkl}P_k P_k$, which originates from the electrostriction coupling. For the $P_3$-component we determine that the term $\frac{F_{11}-F_{44}-F_{12}}{2}Q_{12}\left((P_1^2 + P_2^2)\frac{\partial P_3}{\partial x_3} - P_3 \frac{\partial(P_1^2+P_2^2)}{\partial x_3}\right)$ is present in the Lifshitz invariant (1f). A similar invariant has been discussed in Ref. [31] in the context of incommensurate phases with a defined chirality. This is consistent with our finding of the formation of a flexon induced by the invariant.

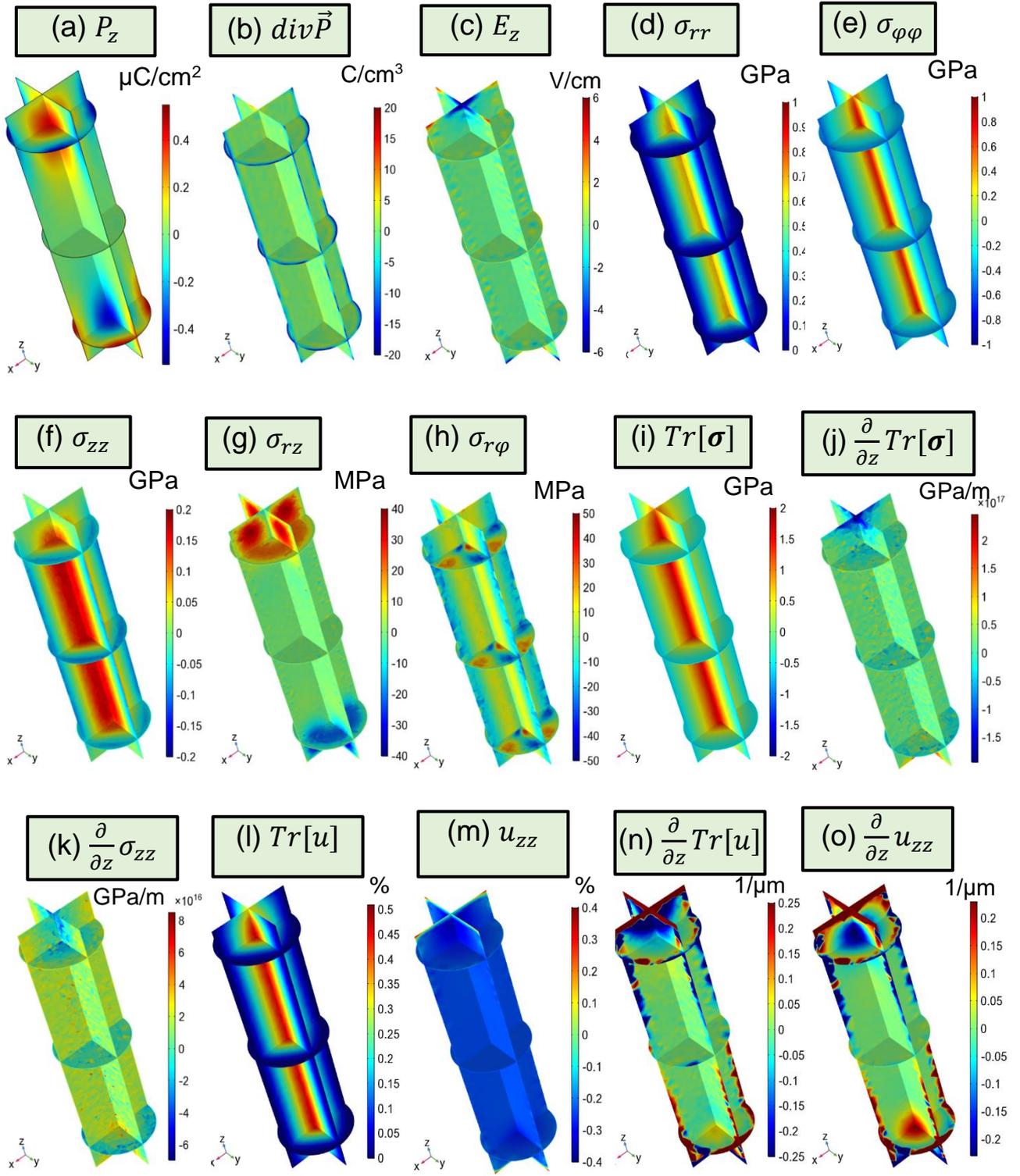

**Figure S9**. Distribution of polarization component $P_z$ (**a**), its divergency $div\vec{P}$ (**b**), electric field component $E_z$ (**c**), nonzero elastic stresses (**d-i**), their derivatives (**j, k**), strain components (**l,m**) and its gradients (**n, o**) in three different cross-sections of the cylindrical core. A cylindrical coordinate system $\{r, \varphi, z\}$ is used for the mechanical stress, strain, electric field and polarization. Note that the stress component $\sigma_{z\varphi} = 0$ (in spherical coordinates). Temperature $T = 300\ K$. All other parameters are listed in **Table SI**.

# APPENDIX F. Topological Index and Toroidal Moment

First, let us calculate the topological index $n$ of the **unit** polarization orientation [32]

$$n = \frac{1}{4\pi}\int_S \vec{p}\left[\frac{\partial \vec{p}}{\partial x} \times \frac{\partial \vec{p}}{\partial y}\right] dxdy \qquad (F.1)$$

for zero, positive, and negative flexoelectric tensor coefficients. The integration is performed over the cylinder cross-section, and here $\vec{p} = \frac{\vec{P}}{P}$.

The flexon polarization has the following structure in Cartesian coordinates

$$P_1 \approx p(r,z)\cos[\varphi(\alpha,z)], \quad P_2 \approx p(r,z)\sin[\varphi(\alpha,z)], \quad P_3 \approx \eta(r,z). \qquad (F.2a)$$

Here $p(r,z) > 0$, and $\varphi(\alpha,z) \approx \alpha - \frac{\pi}{2}$, where $\alpha$ is the polar angle, $x = r\cos\alpha$, and $y = r\sin\alpha$. The polarization magnitude is $P(r,z) = \sqrt{p^2(r,z) + \eta^2(r,z)}$.

Let us introduce the unit polarization as:

$$\vec{p} \approx \{\sin[\theta(r,z)]\cos[\varphi(\alpha,z)], \sin[\theta(r,z)]\sin[\varphi(\alpha,z)], \cos[\theta(r,z)]\}, \qquad (F.2b)$$

where the functions $\sin[\theta(r,z)] = \frac{p(r,z)}{\sqrt{p^2(r,z)+\eta^2(r,z)}}$ and $\cos[\theta(r,z)] = \frac{\eta(r,z)}{\sqrt{p^2(r,z)+\eta^2(r,z)}}$.

The gradients of the unit polarization can be written as

$$\frac{\partial \vec{p}}{\partial x} \approx \left\{\cos[\theta]\cos[\varphi]\frac{\partial \theta}{\partial r}\frac{x}{r} - \sin[\theta]\sin[\varphi]\frac{\partial \varphi}{\partial \alpha}\frac{\partial \alpha}{\partial x}, \cos[\theta]\sin[\varphi]\frac{\partial \theta}{\partial r}\frac{x}{r} + \right.$$
$$\left. + \sin[\theta]\cos[\varphi]\frac{\partial \varphi}{\partial \alpha}\frac{\partial \alpha}{\partial x}, -\sin[\theta]\frac{\partial \theta}{\partial r}\frac{x}{r}\right\}, \qquad (F.2c)$$

$$\frac{\partial \vec{p}}{\partial y} \approx \left\{\cos[\theta]\cos[\varphi]\frac{\partial \theta}{\partial r}\frac{y}{r} - \sin[\theta]\sin[\varphi]\frac{\partial \varphi}{\partial \alpha}\frac{\partial \alpha}{\partial y}, \cos[\theta]\sin[\varphi]\frac{\partial \theta}{\partial r}\frac{y}{r} + \right.$$
$$\left. + \sin[\theta]\cos[\varphi]\frac{\partial \varphi}{\partial \alpha}\frac{\partial \alpha}{\partial y}, -\sin[\theta]\frac{\partial \theta}{\partial r}\frac{y}{r}\right\}, \qquad (F.2d)$$

Here we used $\partial r/\partial x = x/r$ and $\partial r/\partial y = y/r$. Below we use the following relations $\frac{y}{r}\cos\alpha = r\sin\alpha\frac{\partial \alpha}{\partial y} \Rightarrow \frac{\partial \alpha}{\partial y} = \frac{x}{r^2}$ and $\frac{\partial r}{\partial x}\sin\alpha + r\frac{\partial \alpha}{\partial x}\cos\alpha \Rightarrow \frac{\partial \alpha}{\partial x} = -\frac{y}{r^2}$. After obvious, but tedious transformations we obtain:

$$\vec{p}\left[\frac{\partial \vec{p}}{\partial x} \times \frac{\partial \vec{p}}{\partial y}\right] = \sin[\theta]\frac{\partial \theta}{\partial r}\frac{\partial \varphi}{\partial \alpha}\left(\frac{x}{r}\frac{\partial \alpha}{\partial y} - \frac{y}{r}\frac{\partial \alpha}{\partial x}\right) \equiv \frac{\sin[\theta]}{r}\frac{\partial \theta}{\partial r}\frac{\partial \varphi}{\partial \alpha}. \qquad (F.2e)$$

Note, that this relation remains unchanged even if one could take into account the radial dependence of polarization magnitude, $P(r,z) = \sqrt{P_1^2 + P_2^2 + P_3^2}$.

Using the representation (F.2b) and reproducing the detailed calculations in Ref. [33], we obtain:

$$n(z) = \int_0^R \sin[\theta(r,z)]\frac{\partial \theta(r,z)}{\partial r}dr \int_0^{2\pi} \frac{\partial \varphi(\alpha,z)}{4\pi \partial \alpha}d\alpha = \frac{1}{4\pi}[\cos[\theta(r,z)]]_{\theta(r=0,z)}^{\theta(r=R,z)}[\varphi(\alpha,z)]_{\alpha=0}^{\alpha=2\pi} \qquad (F.3a)$$

Here $R$ is the cylinder radius. Substituting here $\varphi(\alpha,z) \approx \alpha - \frac{\pi}{2}$ and $\cos[\theta(r,z)] = \frac{\eta(r,z)}{\sqrt{p^2(r,z)+\eta^2(r,z)}}$ we obtain

$$n(z) \approx \frac{1}{2}\left(\frac{\eta(R,z)}{\sqrt{p^2(R,z)+\eta^2(R,z)}} - \frac{\eta(0,z)}{\sqrt{p^2(0,z)+\eta^2(0,z)}}\right) = -\frac{\eta(0,z)}{2\sqrt{p^2(0,z)+\eta^2(0,z)}},\quad\text{(F.3b)}$$

since $\eta(R,z) = 0$ and $p(R,z) > 0$.

The dependence $n(z)$ is shown in **Fig. S10a** for zero (green horizontal line), positive (3 solid curves), and negative (3 dashed curves) flexoelectric tensor coefficients $F_{ij}$. The black curves $F_{ij}$ values are listed in **Table SI**; the red curves are calculated for twice the value of $F_{ij}$ (labeled as "$2F_{ij}$") and the blue curves are calculated for six times the value of $F_{ij}$ (labeled as "$6F_{ij}$"). The Z-profile of the axial polarization $P_3(0,z)$ and polarization magnitude $P(0,z)$ are shown in **Fig. S10b** and **Fig. S10c**, respectively. Symbols are calculated by FEM for positive $F_{ij}$ (black diamonds), $2F_{ij}$ (red triangles) and $6F_{ij}$ (blue squires). An applied voltage is absent in **Fig. S10**.

Solid and dashed curves are the interpolation functions.

$$P_3(0,z) = f\frac{z}{L}(1 + Az^2)\left(\tanh\left[\frac{z+z_m}{z_0}\right] - \tanh\left[\frac{z-z_m}{z_0}\right]\right),\quad\text{(F.4a)}$$

$$P(0,z) = g\frac{1}{L}\sqrt{1 + \frac{z^2}{B}}(1 + Az^2)\left(\tanh\left[\frac{z+z_m}{z_0}\right] - \tanh\left[\frac{z-z_m}{z_0}\right]\right),\quad\text{(F.4b)}$$

$$n(z) = -\frac{f}{g}\frac{z}{\sqrt{1+(z^2/B)}}.\quad\text{(F.4c)}$$

Here $f$, $g$, $A$, $B$, $z_0$, and $z_m$ are the fitting parameters to FEM results, which are listed in **Table SII**. The length scale $L = 1$ nm. The amplitude $f$ increases with the increase of flexoelectric coupling strength and saturates at high $|F_{ij}|$. Since the value $P(0,z)$ is very close to the $P_3(0,z)$ near the cylinder end, but $P_3(0,z)$ vanishes in the nanoparticle center, the topological index continuously changes from -½ to +½ with a z-coordinate change from one cylinder end ($z = -20$ nm) to another ($z = +20$ nm).

**Table SII**. Fitting parameters for Eqs.(F.4)

| Fitting parameter | Flexoelectric coefficients $F_{ij}$ | | | | | | |
|---|---|---|---|---|---|---|---|
| | $-6F_{ij}$ | $-2F_{ij}$ | $-F_{ij}$ | 0 | $F_{ij}$ | $2F_{ij}$ | $6F_{ij}$ |
| $f$ (μC/cm²) | –0.011 | –0.0071 | –0.0045 | 0 | 0.0045 | 0.0071 | 0.011 |
| $g$ (μC/cm²) | 0.0305 | 0.0224 | 0.0174 | N/A | 0.0174 | 0.0224 | 0.0305 |
| $A$ (nm⁻²) | 0.0100 | 0.0125 | 0.0110 | N/A | 0.0110 | 0.0125 | 0.0100 |
| $B$ (nm²) | 5 | 10 | 15 | N/A | 15 | 10 | 5 |
| $z_0$ (nm) | 1.8 | 2.2 | 2.3 | N/A | 2.3 | 2.2 | 1.8 |
| $z_m$ (nm) | 17.8 | 17.5 | 17.5 | N/A | 17.5 | 17.5 | 17.8 |

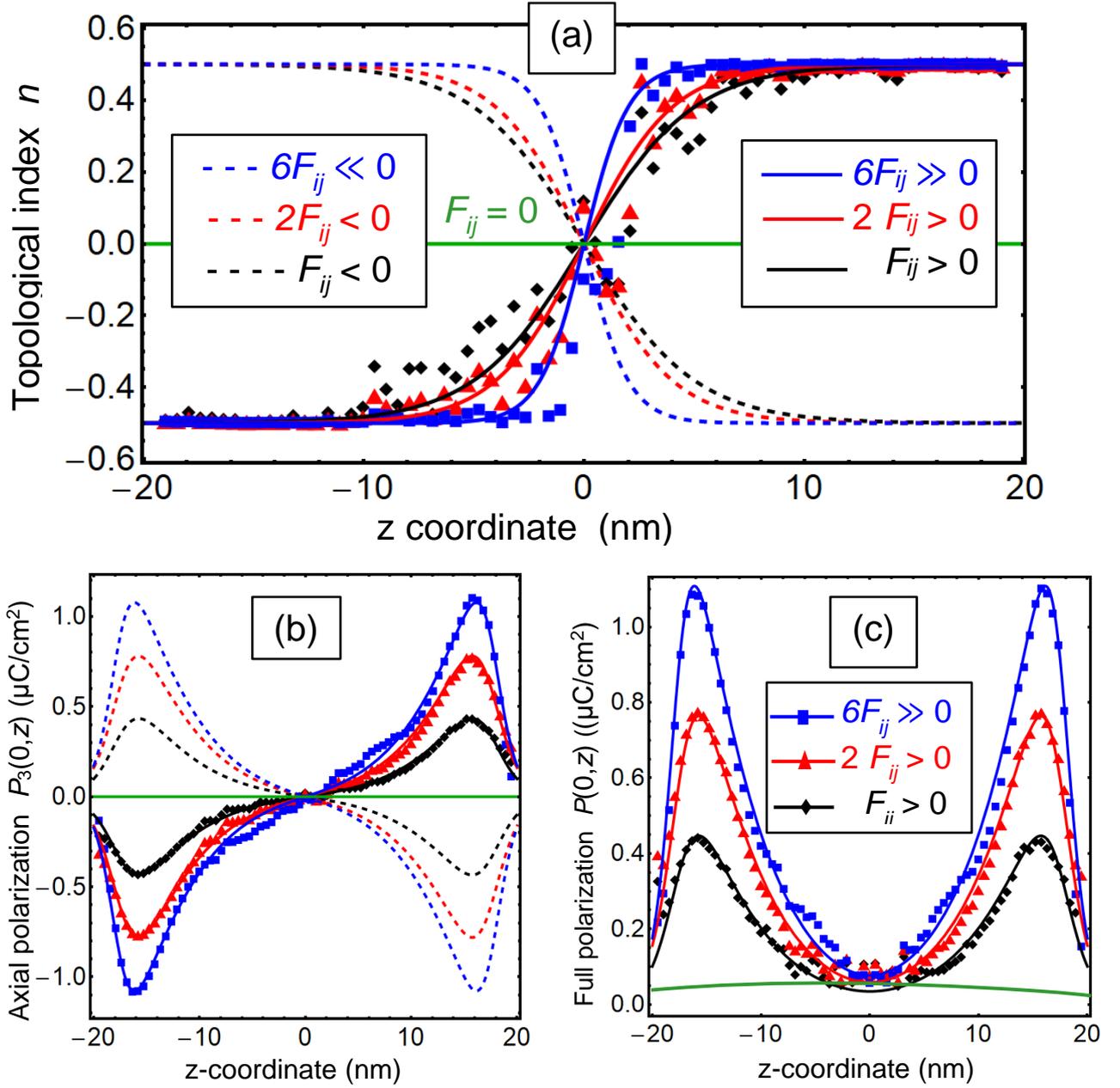

**Figure S10.** (a) Z-dependence of the polarization orientation topological index $n(z)$ for zero (green horizontal line), positive (solid curves), and negative (dashed curves) flexoelectric tensor coefficients $F_{ij}$. For the black curves $F_{ij}$ values are listed in **Table S1**, for the red curves we used $2F_{ij}$, and blue curves, we used $6F_{ij}$. (b-c) Z-profile of the axial polarization $P_3$ (b) and polarization magnitude $P$ (c) calculated at $r = 0$. Symbols are calculated by FEM for positive $F_{ij}$ (black diamonds), $2F_{ij}$ (red triangles), and $6F_{ij}$ (blue squires). Solid and dashed curves are fitting functions listed in the text. Referenced values of $F_{ij}$ and all other parameters are given in **Table SI.** Temperature $T = 300\,K$, $U = 0$.

As a next step, let us calculate the toroidal moment (**TM**):

$$\vec{M} = \frac{1}{V}\int_V [\vec{P} \times \vec{r}]d^3r \qquad (F.5)$$

The integration here is performed over the nanoparticle volume $V = \pi R^2 h$. The toroidal moment (F.5) is independent on the coordinate origin of radius-vector $\vec{r}$, only if all three components of electric polarization have a zero average over the nanoparticle volume. This is true in the considered case for $U = 0$ (no applied voltage).

Let us estimate the toroidal moment using the following approximation for polarization distribution:

$$P_1 \approx p(r,z)\sin[\alpha], \quad P_2 \approx -p(r,z)\cos[\alpha], \quad P_3 \approx \eta(r,z), \qquad \text{(F.6a)}$$

where the coordinates are

$$x = r\cos[\alpha], \qquad y = r\sin[\alpha], \qquad z = z. \qquad \text{(F.6a)}$$

The vectorial product is:

$$[\vec{P}\times\vec{r}] = -\vec{e_x}(p(r,z)\cos[\alpha]\,z + \eta(r,z)r\sin[\alpha]) + \vec{e_y}(\eta(r,z)r\cos[\alpha] - p(r,z)\sin[\alpha]\,z) +$$
$$\vec{e_z}p(r,z)r \qquad \text{(F.7a)}$$

After the integration we obtain that only the z-component of the TM is nonzero:

$$\vec{M} \approx \vec{e_z}\frac{2}{R^2}\int_0^R p(r,z)r\,dr. \qquad \text{(F.7b)}$$

Since the magnitude $p(r,z)$ is almost independent of the flexoelectric coupling (see e.g. **Fig. S5,** the middle row), the TM appears nearly the same for zero, positive, and negative flexoelectric tensor coefficients. To make an analytical estimate in Eq.(F.7b), one can use the following approximation for the magnitude $p(r,z) \cong p_0 \tanh\left(\frac{r}{r_0}\right)$, where the $p_0$ and $r_0$ are temperature-dependent. This results in:

$$\vec{M} \approx \vec{e_z}p_0\left\{1 - d^2\left(\frac{\pi^2}{12} + \text{PolyLog}\left[2, -e^{-\frac{2}{d}}\right]\right) + 2d\text{Log}\left[1 + e^{-\frac{2}{d}}\right]\right\}, \qquad \text{(F.8)}$$

where $d = \frac{r_0}{R}$ and $\text{PolyLog}[x]$ is a polylogarithmic function. **Figure S11** shows the dependence of the normalized TM $\frac{M}{p_0}$ on the parameter $d$.

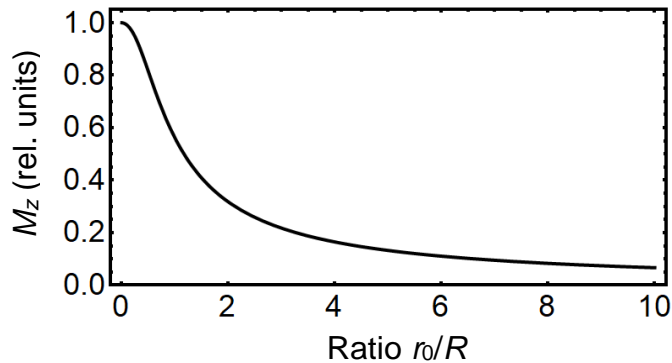

**Figure S11.** The dependence of the normalized toroidal moment $\frac{M}{p_0}$ on the parameter $d$.

In contrast to the topological index, the toroidal moment, $\vec{M} = \frac{1}{V}\int_V [\vec{P} \times \vec{r}] d^3r$, appears almost the same for zero, positive, and negative flexoelectric tensor coefficients. The reason $\vec{M}$ is, for the most part, unaffected by the flexoelectric effect is that $\vec{M}$ is equal to the integral of polarization magnitude $p(\rho, z)$, namely $\vec{M} \approx \vec{e_z}\frac{2}{R^2}\int_0^R p(\rho, z)\rho d\rho$, where the magnitude $p(\rho, z)$ is nearly independent on the flexoelectric coupling (e.g., **Fig. S5,** the middle row).